\documentstyle{article}

\title{Black Hole Thermodynamics in Semi-Classical and Superstring Theory}
\author{Sascha Vongehr}
\date{27 August, 1997}
\begin{document}
\maketitle

\begin{abstract}
This is a revised and shortened version of a MSc thesis submitted to
the University of Sussex, UK.
 
An introduction into the pre-string physics of black holes and related
thermodynamics is given. Then, starting with an introduction of how
superstring theory is approaching the problem of black hole entropy,
work on that and closely related topics like
Hawking radiation and the information paradox is reviewed.\\
\end{abstract}

\tableofcontents
\newpage

\part{Introduction}
The aim of this thesis is to give an introduction to the subjects of
black holes and related superstring theory (but just to the extent of
facilitating the understanding of this review)
and then to review the progress of how far superstring theory has got in
providing the statistical mechanics of black holes.  This is an intriguing subject
for anyone interested in fundamental sciences because two very
important but quite different theories (general relativity and a form of
quantum mechanics, here
quantised superstring theory) meet each other at this point and moreover do so at their
thermodynamical ends. In a sense, thermodynamics is still the most fundamental
theory of physics (with statistical mechanics only coming in when based
on information theory) because physics is based on observations
(perception of macroscopic entities) that have to be irreversible in order
to constitute a measurement. However, most advances in
thermodynamics have been
made by taking statistical mechanics as the underlying
theory and thermodynamics as its surface phenomenon -- popping up with
overwhelming likelihood. Superstring theory recently succeded in providing 
models for the microstructure of black holes. Entropies, temperatures and
black body radiation of certain simple black holes have been calculated and shown
to be equivalent to the same variables of those black holes in the context of
classical general relativity. Moreover, sparked by the Strominger-Vafa computation
~\cite{S+V96} many people shortly after followed with calculations for
quite a variety of black holes in different numbers of dimensions of spacetime
with charges and even angular momentum and often the black hole solution in the
context of classical general relativity has not been known before the stringy approach
succeded.
  
The key points of these calculations are the identification of soliton
black holes in superstring theory with configurations of so called
D-branes (to be introduced later on). Full validity of the
computations applies for weak coupling where the string length is larger
than the Schwarzschild radius. For a black hole we require that the
string length sits inside the event horizon and that means strong coupling. One
argues that the calculations are still valid for extremal black holes 
because the BPS (Bogomol'nyi-Prasad-Sommerfeld)
properties of the D-branes make the results coupling independent, i.e.
protected by supersymmetry.

Superstring theory's low energy effective actions are supergravity
actions and many methods of general relativity can be applied straight
forwardly to supergravity. Therefore, a relatively long introduction to
classical and semi-classical black hole physics is included in
order to show important methods only referred to later on and in order
to provide insight into the fact that black hole solutions are not just any
ol' examples of a theory because via these solutions one introduces arrows
of  time (at the event horizon) for instance and thermodynamics in
general. ``One could say that they are
the `Hydrogen atom' of quantum gravity.'' ~\cite{Malda96}
 
It might surprise how much (extensively) but at the same time as well how 
little supergravity
one really needs (to know, intensively) in order to follow the
argumentations. The assumed background knowledge is bosonic string theory and
relativistic quantum field theory but only
rudimentary superstring theory and general relativity are required.
Often all constants like $ \hbar ,c,G,k_{B} $ will be suppressed and are
only written
in case they give a deeper understanding when thought of as variables.

\newpage
\part{The Pre-String Physics of Black Holes}
\section{Introduction}
\subsection{General Relativity}
General relativity is a very geometrical and topology involving theory
but the connection between the clearly understood geometry on one side
and its dependence on the energy distribution on the other (of equation
(\ref{RT}) for example) is not well understood at all. The more work is
done in order to resolve this connection the more it becomes apparent
that the description of spacetime as 3+1 dimensional instead of 2+1 or 9+1
or whatever is only the easiest choice when doing mesoscopic physics. It
is a surface phenomenon. General relativity is an effective low energy
theory about the elasticity of spacetime, i.e. its resistance against
curving it. That is where inertia (mass) is coming from in this model
and I think it has a rather nice parallel in special relativity where
acceleration has the units of curvature $(1/length)$ and inertia is therefore the
resistance a body puts against curving its worldline (that is in proper
natural or geometrized units using length to measure time). Viewing
general relativity as a mesoscopic theory means that it might be not so
fundamental after all and that quantising it could be like quantising
phonons ~\cite{Hu96}. ~\cite{Hu96} put forward the idea that general
relativity is the ``geometro-hydrodynamics'' of a microscopic theory. 

Many people are taught that relativity is ``very relativistic'' in the
sense that any reference to absolute media equals a failure of the right
understanding. One of the reasons is that special relativity has a
derivation on an operational basis (how do we measure). Today, I think
that special relativity is easiest understood as being the physics that
systems would encounter if they were out of waves (on a pond of fluid)
with maximal wave velocity $c$. With no way that these systems of bound
waves would ever perceive the fluid in low energy experimentation (no
sound in, splashing of the fluid) they will find time dilatation, length
contraction and so on (but no way to tell which system is stationary in the fluid
reference system). Even in \emph{general} relativity there are aspects
of this. A region where such a medium is sucked in (a sink) will have an
event horizon where the medium has velocity $v=c$ into the sink and even
the fastest waves cannot get outside anymore (Of course --- a fluid
giving rise to \emph{general} relativity needs very strange properties). 
I am not aware of anybody having proposed this fluid toy model 
but I would not be surprised if so and particle or condensed state physicists
will surely find it to be an obvious ``relativation'' to see relativity like this. 

Applying Ockham's razor\footnote{``Ockham's razor'' frees theories from
variables that are not observable in principle. It is certainly rightly
applied to the idea of hidden variables in quantum theory.}and refusing
the possibility of an absolute medium or reference system (like
the mean microwave background) adds nothing to relativistic theories
because they can be described in any reference system and thus even in the
absolute one. It only removes a possible connection to more fundamental
physics. By the way, \emph{general} relativity has not got an
operational derivation like special relativity has.  All this is just to prepare
those readers who were introduced to relativity theories as being ``very
fundamental'' and ``very relativistic'' for a more open point of view. 

One should mention though that this (general relativity as an effective
theory) is not the nowadays and everywhere accepted position. There are
suggestions that general relativity is more fundamental than quantum
mechanics and one might expect this because the former is non-linear
and the latter is linear (e.g. ~\cite{Ash97} and references therein).
Quantum physics would be based on the topological nature of the
interplay between measured system and the boundary conditions set by the
measurement apparatus ~\cite{Hadley97}.

In general relativity a solution means an expression that solves the
famous Einstein field
equations:
\begin{eqnarray}
R^{ \mu \nu } - \frac{1}{2} R g^{\mu \nu} = (8 \pi ) T^{\mu \nu }
\label{RT}
\end{eqnarray}
where $T^{\mu \nu}$ is the energy-momentum tensor, $R^{\mu \nu}_{(g^{\mu \nu})}$ the
Ricci tensor and $R$ the Ricci curvature scalar $R = R^{\mu \nu} g_{\mu
\nu}$ (sometimes one sees different sign conventions). Thus a solution
means an expression for the metric tensor $g_{\mu \nu}$  and  the
energy-momentum tensor; note that the $R$-variables above dependent only on the
metric alone.

The action for the general relativity that includes only gravitating
matter/energy  and electromagnetic fields $(F=dA)$ is 
\begin{eqnarray}
S \propto \int \, d^{4} x \, \sqrt{-g} \, (R- \frac{1}{4} F^{2} ) \label{action}
\end{eqnarray} 
where $ g = det \,  g_{\mu \nu }$. 

A vacuum solution with no radiation and no electro-magnetic fields      
gives the metric for $T^{\mu \nu} = 0$ everywhere which leads with the
$R^{\mu \nu} \leftrightarrow (8 \pi ) T^{\mu \nu}$-duality of the
Einstein equations (\ref{RT}) to $R^{\mu \nu} = 0$. Thus a vacuum
solution is an expression
for $g_{\mu \nu}$  dependent on the coordinates that solves:
\begin{eqnarray}
R^{\mu \nu}_{(g^{\mu \nu})} = 0
\end{eqnarray} 

All what we need to know here is that doing general relativity requires
that the flat Minkowski metric
\begin{eqnarray}
 \eta_{\mu \nu} = diag \, (1,-1,-1,-1)
\end{eqnarray}
is replaced by the metric matrix $g_{\mu \nu}$ that can be far more
complicated depending
on the matter/energy distribution or even pure empty space curvature. 
$ g_{\mu \nu} =  \eta_{\mu \nu} $
holds only for a flat spacetime where then special relativity is valid
again. For many physicists, learning about the Einstein equations is a
``one off'' only to realise that they are far too complicated to be of
much use to them.  From 
then onwards, all they need is an ansatz for the metric that
reflects the symmetry of the system (e.g.: spherically symmetric).
Trying to obtain $ g_{\mu \nu} $ starting with $ T_{\mu \nu} $ is futile because
$ T^{\mu \nu} $ depends on $ g^{\mu \nu} $ as well.
The metric is there to calculate distances (define a geometry) 
and leads therefore together with the coordinates one chooses to the
line element $(d\tau )^2 =  g^{\mu \nu }_{(x)} \, (dx_{\mu }) \, (dx_{\nu }) $
 that gives the physical
distances $ \tau = \int \, d \tau $ due to paths one goes in pure
coordinate space. I write $\tau$ instead of $s$ because it is actually the eigentime, the
time a clock would proceed if its worldline in spacetime were the path
choosen (consider the sign convention for $\eta ^{\mu \nu}$).\\

General relativity is easily generalized to $d$ spacetime dimensions by
writing down a $d \times d$-metric tensor. To investigate general
relativity in other that $d=3+1$ dimensions can have advantages.
\begin{itemize} 
\item $d=2+1$ general
relativity for example is shown to be equivalent to the Yang-Mills
theory of a Chern-Simons action ~\cite{Witten88} ~\cite{Bimonte97}.
Chern-Simons terms
are Lagrangians like:
\begin{eqnarray}
 L = (m/4) \epsilon ^{\mu \nu \rho} F_{\mu \nu } A_{\rho } \, \, or \,
\,
 L = (m/4) \epsilon ^{\mu \nu \rho} A_{\mu} (\partial _{\nu} A_{\rho} )
\end{eqnarray}
where m is the Chern-Simons mass of the gauge field.
One can then approach general relativity from this Yang-Mills point of
view more 
easily than in $3+1$ dimensions because we cannot do the same there ---
that is, a pure Yang-Mills description.

\item A $d=1+1$ gravity is the $d=2+1$ gravity with imposed axial
symmetry. This yields Jackiw-Teitelboim dilaton gravity ~\cite{Jack84}
where the
dilaton is related to the length of the path around the axis of
symmetry at any given point. It is called ``dilaton gravity'' because
the dilaton is not just minimally coupled\footnote{Minimal substitution
is one procedure of introducing fields to relativity theory that are not
in the theory from the outset.} This theory is interesting in
conection with our subject because it can be described by the
sine-Gordon theory and the black holes are then solitons of the sine-Gordon
system (on sine-Gordon theory and solitons see ~\cite{Raj82}). It might   
be that the entropy of those black holes can be accounted for by
configurations of the members of the so called ``breather solutions''
(bound solitons) ~\cite{Gegenberg97}.

\item In order to compare $D$
dimensional superstring theory with general relativity one often needs 
more than $d=3+1$ dimensions because the superstring theory can have less
than $D-4$ dimensions compactified. I use $D$ for the dimensionalty of
the full theory and $d<D$ for the dimensionality that a large scale
observer will see. 
\end{itemize}

Masses (and in general other momenta apart from $p^{0} = mass$ ) we
can calculate from the metric with the ADM-Method
(Arnovitt Dazer Mizner) ~\cite{Weinberg}. One uses the metric at
 $r \rightarrow \infty$ where
one can use the weak field approximation:
\begin{eqnarray}
g_{\mu \nu} = \eta _{\mu \nu} + h_{\mu \nu} \label{gexpans} \\
\lim_{r \rightarrow \infty} h_{\mu \nu} = 0 \label{weakfield}
\end{eqnarray}
Since $g_{\mu \nu}$ solves the Einstein equations there exists a
conserved energy-momentum vector $ p^{\mu } $. With $ \eta_{\mu \nu } =
\eta_{\mu } \eta_{\nu } $ etc. and for $\mu = 0$ :
\begin{eqnarray}
h_{i} = ( \frac{\partial h_{jj}}{\partial x^{i}} -
 \frac{\partial h_{ij}}{\partial x^{j}}) \\
p^{0}  = (- \frac{1}{16 \pi G}) \int r^2 \, d\Omega \, \eta_{i}
(h_{i})_{\infty}
\end{eqnarray}
An example is given in the next section.

\newpage
\subsection{Kerr-Neumann Solutions and Extremal Black Holes}
It is sufficient for our purposes to define a black hole as a region of no
escape. Neither matter nor light can ever leave it. For more
sophisticated physics this is not too good a definition because any
future light cone in flat Minkowski space or the universe itself for
example is such a region. However, definitions like this allow us to
identify black holes when they emerge in a theory since the (spatial) boundary
of a region of no escape is the event horizon where even light is accelerated
inwards enough to make it stand still relative to spatial infinity and
thus the event horizon often reveals itself as a coordinate singularity of the
space-time metric that can be circumvented by transforming to a new
coordinate system. 

The stationary mass-only solution, that is the uncharged and
 not rotating black hole called
Schwarzschild solution, is a vacuum solution (apart from the singularity in
the middle of it). Thus, a Schwarzschild black hole is a vacuum solution that
has an event horizon and the latter revealed itself as a coordinate singularity of the
metric without it being a curvature singularity.

All classically stationary black holes fall into the 
3-parameter family of Kerr-Neumann solutions where the parameters are
mass M, angular momentum density a=J/M and electric charge q. More
precisely, this holds only for Einstein-Maxwell black holes, that is for
ones that have no self-gravitating Yang-Mills fields for example. The
proofs of general relativity are mathematically demanding but physically
the results are often expected
if not trivial statements.  That a stationary black hole is axissymmetrical for
example is expected since otherwise tidal forces will lead to
gravitational radiation and the black hole loses energy
-- hence is not stationary.  That there are only three parameters is due
to the fact that for gravity and self-gravitating Maxwell fields no multipole moments
can  be seen outside the black hole
leaving only overall values of long range fields as characteristics
although inside the black hole multipol momenta may exist -- even off
centre~\cite{R.Wald.72}. That only those parameters survive that can be
calculated via surface integrals at spatial infinity is called the ``no hair
theorem'' ~\cite{Israel67} ~\cite{Carter71} ~\cite{Haw72} . Any
properties that cannot be detected at spatial infinity but can be
detected close to the black hole
are refered to as ``hair'' of the black hole. It will occur for any
non-linear theory that is coupled to gravity ~\cite{Nun96} like in case
of self-gravitating Yang-Mills fields for which no uniqueness theorem
like the Kerr-Neumann family exist. 

The line element of the Kerr-Neumann solution is
\begin{eqnarray}
(d\tau )^2 & =\frac{\Delta -a^2\sin ^2\Theta }{\Sigma } (dx^0)^2+\frac{2a\sin ^2\Theta
(r^2+a^2-\Delta )}{\Sigma } dx^0 d\phi \nonumber \\
 & - \frac{\Sigma }{\Delta } (dr)^2 - \Sigma (d \Theta )^2-
\frac{(r^2+a^2)^{2}
-\Delta a^2\sin ^{2}
\Theta }{\Sigma }\sin ^2\Theta (d \phi )^2 \label{lineelement} \\
 & \Sigma=r^2+a^2\cos ^2\Theta \\
 & \Delta = r^2 + a^2 + q^2 - 2Mr
\end{eqnarray}
\cite{R.Wald.84}, where it can be very misleading to interpret the coordinates
$r, \phi $ and $\Theta $ as spherical polar ones and $x^0 $ as the time. At large $r$
this interpretation is not misleading since $g_{\mu \nu} =  \eta_{\mu
\nu}$ there and the line element is just the one of special relativity
(here in spherical polar coordinates). Far from the centre $r=0$ of the black
hole the spacetime is flat. But for the black hole (i.e. the very region of no
escape behind the event horizon) the line element
gives timelike results if going along $r$ and spacelike ones for lengths
along $x^0 $. This mirrors the fact that nothing can escape, i.e. that 
the future light cones are entirely inside the black hole.    

M is the mass of the black hole and q will be
generalised later on to include new kinds of charges. $(q=0,a\neq 0)$ is the Kerr
black hole, 
$(q\neq 0,a=0)$ is the Reissner-Nordstr\"{o}m solution and $(q=a=0)$ the
famous Schwarzschild solution.  While $(\Sigma =0,M\neq 0)$ is the true
curvature singularity the $(\Delta =0)$ -singularities are due to the
choice of coordinates and are in fact the event horizons at 
\begin{eqnarray}
r_{\pm}=M\pm \sqrt{M^2-(q^2+a^2)}
\end{eqnarray}
where $r_{+}$ is the outer event horizon that we will be concerned with
mostly. These horizons only exist for $(q^2 + a^2 \leq M^2)$. $(q^2+a^2=M^2)$ are
the extremal black holes.
The extremal black holes are not so much the big astrophysical objects
because a
$(q^2 + a^2 = M^2)$-solution must be very small in order not to attract
opposite charge immediatelly in all astronomically realised situations.
It is therefore, if at all, a microscopic
one. It is an infinitesimal step close to a naked singularity and
because the event horizon and singularity coincide it is called
``null-singularity''.

For $(q^2+a^2>M^2)$
the curvature singularities are naked (without event horizon
because $r_{\pm }$ would be imaginary then). A short calculation shows
that general relativity naively \footnote{The validity
of general relativity in the microcosmos and so close to elementary particles
has always been highly doubtful of course.} applied to an electron results in a
naked singularity and not in a black hole. Just
take known values for the charge and the spin of the electron and
compare them with its mass. On both accounts alone (charge or spin) I
found that $(q^2 + a^2 )$ is far greater than $M^2 $ (by a factor of
$10^{100} $ for the charge in SI units). Nevertheless, we
will be confronted with extremal black holes later on and many proofs, for example
the one proving that every black hole has a curvature singularity, use the
cosmic-censorship conjecture, which is basically saying that there are no
naked singularities possible or better none but the one of the big bang. The
cosmic censorship conjecture is widely doubted. Either people with a
background in general relativity are not afraid of naked singularities
as they are used to strange geometrical objects by now, or people
with mostly a background of particle physics deny the singularity in the
first place.  

Note that the solutions have a well behaved metric even for negative
$r$. In general, there are many transformations of the coordinates
possible that seem to open up new regions of spacetime. These regions
exist ``beyond'' the singularity $ (r<0) $ and are asymptotically flat
just like the solution is at large values of positive $r$. Whether or
not there are whole universes behind the horizons and beyond the singularities
is a matter of debate. Coordinates are names for points like
temperatures are names for states of thermal equilibrium. There are not
necessarily physical and accessible regions corresponding to previously
not considered ranges of names. In the context of general relativity
coordinates are easily misinterpreted anyway.   

Recall that one is free to generalise general relativity to $d$ spacetime
dimensions. For the Reissner-Nordstr\"{o}m solution for example this leads to: 
\begin{eqnarray}
(d \tau )^{2} = \frac{\Delta }{\Sigma } (dx^{0})^{2} - \frac{\Sigma }{\Delta }
(dr)^{2} - r^{2} \, d\Omega ^{2}_{d-2} \\
\frac{\Delta }{\Sigma } = (1- \frac{r_{+}^{d-3}}{r^{d-3}}) 
(1- \frac{r_{-}^{d-3}}{r^{d-3}}) \label{lineelementd} 
\end{eqnarray}
where $r_{+} = r_{-}$ is the extremal black hole $(M=q)$. It is easy to
check that for the mass and the charge one obtains the following
arithmetic and harmonic means:
\begin{eqnarray}
G^{d}_{N} M = \frac{r_{+}^{d-3} + r_{-}^{d-3}}{2} \times const_{(d)} \\
G^{d}_{N} q = \sqrt{r_{+}^{d-3}  r_{-}^{d-3}} \times const_{(d)}
\end{eqnarray}
where I restituted the gravitational (Newton) constant $G_{N}^{(d)}$ of
units $mass^{2-d}$ and 
where\\  $const_{(d=3+1)} = 1 $ and  $const_{(d=4+1)} = 3 \pi / 4 $ for
example derive from the volumes of spheres of higher dimensionality. 
This leads for the line element of the Schwarzschild solution (no charge, no
angular momentum) for instance in $d$ dimensions to:
\begin{eqnarray}
\frac{\Delta }{\Sigma } = (1-\frac{2G_{N}M}{r}) \rightarrow
(1-\frac{2G_{N}^{(d)}M}{const_{(d)} \, r^{d-3}})   \label{deltaoversigma}
\end{eqnarray}

The calculation of the masses of the black hole solutions can be done
with the ADM method (equations (\ref{gexpans} ), (\ref{weakfield} ) \ldots ). For the
 Schwarzschild solution for instance in Cartesian coordinates:
\begin{eqnarray}
(d\tau)^2 & = (1-\frac{2G_{N}M}{r})(dt)^2 - \, dx^i \,dx^j [\delta _{ij}
+\frac{2G_{N}Mx^i x^j}{r^2 (r-2G_{N}M)}]\\
 & h_{00}= - \frac{2G_{N}M}{r}\\
 & h_{ij}= - \frac{2G_{N}M x^i x^j}{r^2(r-2G_{N}M)} \rightarrow - 
 \frac{2G_{N}M x^i x^j}{r^3}\\
 & \Rightarrow (h_{i}) _{\infty} = (-4G_{N}M\frac{\eta_{i}}{r^2}) \, \Rightarrow p^0 =M
\end{eqnarray}

\section{Thermodynamics}
\subsection{Massloss, Arrow of Time and 0th to 3rd Laws}
\label{0-3}
Even in the context of classical general relativity (without effects of
quantum mechanics) it is not true
that a black hole cannot lose some of its mass M. The latter is only valid for
the Schwarzschild solution. The black hole as a region of no escape loses mass by taking up
negative energy. The Kerr and the Reissner-Nordstr\"{o}m solutions can lose all their
rotational or electrical field energy.  This is known as the Penrose
process in case of the Kerr solution (see~\cite{R.Wald.84} for an introduction to the
Penrose process). For both, the Kerr and the Reissner-Nordstr\"{o}m black holes, but
not for the charged Kerr black hole, it is possible to consider Maxwell's
linearised Einstein equations in the background of the black hole in order to reduce
the problem of determining the behaviour of electro-magnetic and gravitational
perturbations to ordinary differential equations (see for
example~\cite{Chandra.S.83}).  With this approach one can deduce the
possibility of superradiant scattering~\cite{Starob73} where electro-magnetic or
gravitational radiation is sent into the ergosphere
\begin{eqnarray}
r_+ < r_{ergo} < M+\sqrt{M^2-q^2-a^2\cos ^2\Theta }
\end{eqnarray}
 outside the black hole in such a
way that the scattering leads to a reflected part with higher amplitude
and a transmitted part going into the black hole and carrying negative energy.
However, the irreducible mass, which is in the case of a Kerr black hole
\begin{eqnarray}
M^2_{irr}:= \frac{1}{2}[M^2+(M^4-J^2)^{1/2}]
\end{eqnarray}
cannot (classically) be reduced:
\begin{eqnarray}
M^2=M^2_{irr}+\frac{J^{2}}{4M^2_{irr}}\geq M^2_{irr}
\end{eqnarray}
and the line element (equation (\ref{lineelement}))
shows that the event horizon at $r_{+}$ only ever
becomes larger. Even in case of a maximally rotating black hole $(J=M^2)$ one
can only extract $(1-\frac{1}{\sqrt{2}})\simeq 29\%$ of the mass.  The
area of the event horizon of the general Kerr-Neumann solution is given by
\begin{eqnarray}
A= \int \sqrt{g_{\Theta \Theta } g_{\phi \phi }} \, d\Theta \,
d\phi \\ 
\Rightarrow A_{+} = \int_{r = r_{+}} \sqrt{(\Sigma ) (\frac{(r^2 + a^2 )^2 - \Delta a^2
sin^2 \Theta }{\Sigma } sin^{2} \Theta )}
 \, d\Theta \, d\phi \\ 
= \int (r_{+}^{2} + a^2 ) sin\Theta \, d\Theta \, d\phi 
\end{eqnarray}
because $ \Delta = 0 $ at the event horizon. It follows:
\begin{eqnarray}
A_{+} = 4\pi (r^2_{+}+a^2) = 16 \pi M^2_{irr}
\end{eqnarray}
and thus the area can only increase, too. $A_{+} \neq 4\pi r_{+}^2 $ because the event
horizon rotates for $a \neq 0$ and is then Lorentz contracted relative to
the observer far away.\\

In fact, a more general area theorem
proven by S. Hawking 1971, saying that the area of the event horizon of
 any black hole can
only ever stay the same or increase, leads in the case of two black holes which
merge and, in doing so, radiate energy in form of gravitational waves,
to $(A\geq A_{1}+A_{2})$ with $(1-\frac{1}{\sqrt{2}})\simeq 29\%$ being
again the maximum loss of mass via gravitational radiation.
This indicates fundamental limits of mass
loss leading to the notion of irreducible mass.  The area theorem is
surprising in that it introduces an arrow of time into the theory of 
general relativity of dense matter (although general relativity  is time symmetric and the
black hole  has a white hole as the time symmetrical partner).  The entropy
time arrow has the same direction. Therefore, the area of a black hole is related
to entropy via a relation between the area theorem and the second law of
thermodynamics. This poses no problems for the macroscopic observer for
whom both, general relativity and thermodynamics are very fundamental
theories. But microscopically the second law is not a rigorous consequence of classical
laws. It is rather something that follows from statistical mechanics with
overwhelming probability for systems with a large number of degrees of
freedom. There is no thermodynamics on the scale of an elementary
particle --- only statistics or better stochastics even. 
Are there black holes ?

With the introduction of a Killing field $\chi^{\mu}$
normal to the event horizon one can define a proportionality factor $\kappa$:
\begin{eqnarray}
D^{\mu}(\chi ^{\nu}\chi _{\nu})=:-2\kappa \chi ^{\mu}
\end{eqnarray}
where $D$ is the covariant derivative.
Some calculation ~\cite{Foster95} leads to  
\begin{eqnarray}
\kappa^{2}_{+}= \lim_{r \rightarrow r_{+}}Va 
\end{eqnarray}
where $V=\sqrt{-\chi ^{\mu}\chi _{\mu}}$ with $ a=\sqrt{-a^{\mu}a_{\mu}}; \, \,
a^{\mu}=(\chi ^{\nu}D_{\nu} \chi _{\mu}).$ 

$\kappa _{+} $ is the
surface gravity at the (outer) event horizon --- that is the force that must be exerted to hold a unit
test mass at the event horizon.  $\kappa _{+}$ is not infinite because both -- force and mass
--- are infinite relative to spatial infinity. The result that $\kappa _{+}$
is constant all over the event horizon of
a stationary black hole is similar to the zeroth law of thermodynamics (one
version of it), saying
that the temperature of a body in equilibrium is the same everywhere.
Using Stokes' theorem (i.e. calculating at spatial infinity similar to
the ADM method) one deduces the following for the mass of a black hole:
\begin{eqnarray}
\delta M=\frac{1}{8\pi}\kappa _{+} \, \delta A_{+} + \Omega _{+}\, \delta J \label{firstlaw}
\end{eqnarray}
where $\Omega _{+}$ is a sort of angular velocity of the event horizon:
\begin{eqnarray}
\Omega = \frac{d\phi }{dx^{0} } = - \frac{g_{0\phi} }{g_{\phi \phi } } \\
\Rightarrow \Omega _{+} = \frac{a}{(r_{+}^2 + a^2) }   
\end{eqnarray}
because of $\Delta = 0$ at the event horizon.
Having
$\kappa _{+}$ behaving similar to temperature $T$ and the area of the event horizon
$A_{+} $ similar to an entropy $S$
one can say that this is the first law of thermodynamics with  $\Omega
_{+} \, \delta J$ being the work term $P\, \delta V$.  The third law --
that one cannot reach zero temperature quasistatically -- is reflected
in the fact that one cannot achieve $\kappa _{+} = 0$ quasistatically.
~\cite{Wald74} showed that the closer the state of the black hole is to the
state of an extreme black hole the harder it is to get even closer (by adding
 charge quasistatically for example) in a manner similar to the third
law. Neither Plank's version of the third law, the possible gauging $(\lim_{T\rightarrow 0}
S=0)$, nor Nernst's statement ($(\lim_{T\rightarrow 0} \Delta S=0)$ for
any process occuring near absolute zero) are appropriate.
They come with the assumption of
classical thermodynamics that the ground state is never degenerate.
We see that the extremal black hole
$[(q^2+a^2)=M^2]$ has an event horizon with positive area because $ r_{+}^2 = r_{-}^2 =
M^2 $ leads to $ A_{-} = A_{+} = 4 \pi (M^2 + a^2 )$
although it is infinitesimally
close to having no event horizon. Its surface gravity $\kappa _{+}$ is zero. The zero
temperature of extremal black holes having finite entropy indicates high
degeneracy.

\newpage
\subsection{The Black Hole is a Black/Grey Body}
Still -- we cannot simply set temperature proportional to $\kappa _{+} $ and
entropy proportional to the area of the event horizon since
classically the black hole has no black body radiation and is only in
equilibrium with a zero temperature background. Moreover, we can violate the
second law by lowering a body quasistatically to the event horizon thereby
retrieving all its energy (even its thermal one (!)). In order to resolve
this paradox and in order to give finite temperature to the black holes
one has to leave the purely classical general relativity.
For example one might just claim that because the
lowered body has to be lowered very close to the event horizon
one comes into the realm of superstring theory as one is closer to the
event horizon than the string length.

There is another way which one might anticipate on the grounds alone
that the black body radiation spectrum led Planck to the assumption that
light is quantised although black body radiation follows from classical
laws and is but a $T,V,P$-equilibrium of the electro-magnetic field. 
What is done is to proceed semiclassically. One introduced quantum
effects into the black hole background. 
The demand for unitary evolution in quantum mechanics
 $(<1>=1$--normalisation for the probabilities needs unitary operators)
does not allow for anything that only absorbs but does not emit. 
~\cite{Hawking75} showed that particle creation near the event horizon leads to an
effective emission of particles from the black hole which has
$(T=\frac{\kappa _{+}}{2\pi})$ -black body
characteristic in case of an Schwarzschild black hole
for an observer at infinity which
is in line with the equivalence principle and the result ~\cite{Unruh76} that an
accelerated $(a)$ observer is being immersed in a thermal bath of
particles with temperature (Unruh-temperature):
\begin{eqnarray}
T_{Unruh} =\frac{a}{2\pi}  \label{Unruh}
\end{eqnarray}
The energy emitted is equal to that of a black body with surface area $A_{+}$.
Thus, $\frac{\kappa _{+}}{2\pi}$ is the temperature of the black hole in a direct
physical sense. In SI--units one writes:
\begin{eqnarray}
k_{B} T_{Hawking} = \frac{\hbar \kappa _{+}}{2\pi c} = \frac{\hbar c^3}{8\pi
G_{N} M}  \label{ab}
\end{eqnarray}
 so that
\begin{eqnarray}
\lim_{\hbar \rightarrow 0}T_{H} = 0
\end{eqnarray}
 and it becomes obvious that we are
dealing with a quantum mechanical  effect. Originally this was derived from studies of
linear, free quantised fields propagating on certain black hole backgrounds but
it holds equally well for fields of arbitrary spin in general black hole spacetimes.
In order to calculate the exact expressions
for scattering cross sections and black hole decay rates one needs to consider the
wave equation of the radiation on the black hole background.
This wave equation (for example Klein Gordon equation on the background)
is obtained by substituting the metric of the black hole background into
the covariant Laplacian:
\begin{eqnarray}
\Box = \frac{1}{\sqrt{|g|}} \partial _{\mu } \sqrt{|g|} \, g^{\mu \nu }
\partial
_{\nu }
\end{eqnarray}

Actually, the black hole's Hawking radiation is not striktly coming
with a black body spectrum. This is due to frequency-dependent filtering
because of the potential barriers due to the long range fields outside the
black hole that reflect some
of the radiation back to the black hole. The black body spectrum would
only be seen in case of no such filtering and only by the distant
observer at spatial infinity. The Hawking formula for the black hole's
decay rate is ~\cite{Hawking75}
\begin{eqnarray}
d \Gamma _{Hawking} = \sigma _{abs \, (\omega , q_{i} )}  \rho _{(\frac{\omega
- q_{i} \phi _{i} }{T_{H}} )} \,
\frac{d^{4} k}{(2 \pi )^{4}} \label{GammaHawking} \\
\rho (x) 	= \frac{1}{e^{x} -1} 
 		= \frac{e^{-x}}{1-e^{-x}} \label{rho}
\end{eqnarray}
where $q_{i}$ are the charges of the emitted particle, $\phi _{i}$
the related chemical potentials and $\rho $ is the thermal occupation
factor (here for bosons the ``mean
Planck excitation'' or ``BE-distribution'') and $\sigma $ is the classical
absorption cross section (grey body factor) that would equal $A_{+}$ in case
of an ideal black body spectrum.
The effect of this partial reflection that depends on the frequency of the light   
is that the semi-classical and the string theory black holes look the
same from far away. This is interesting because the filtering occurs
outside the black hole and introduces new complications into the
discussion of whether the information of the black hole is stored near
the singularity or on the event horizon.  

For a very simple primer on Hawking radiation see ~\cite{Open79}, for
an introduction of how to solve the paradox of the second law violation
near an event horizon please see ~\cite{Wald88}.

The Hawking radiation opens up a way 
to the extremal black hole as being the final state of the Hawking
process (effective evaporation of a black hole starting when the background
temperature drops under the temperature of the black hole) of a charged black hole
since the Hawking radiation lowers the classically irreducible mass $M_{irr}$.   
Introducing more charges like the magnetic monopole charge for instance
leads to a lot more possible extremal black holes that satisfy the
$q^2 + a^2 = M^2 $ bound with different contributions of the charges involved.

\subsection{The Calculation of Temperature and Entropy}
Black hole thermodynamics can be approached by using Euclidean
path integrals ~\cite{Gibb77}. This method uses the
Einstein-Hilbert action for gravity and actions for classical matter
fields. Time is analytically continuated into the complex numbers. One
formulates the path-integral after a Wick rotation $(x^0 \rightarrow i\tau)$. 
The action is expanded semi-classically via a saddle point expansion around
classical field configurations. One finds partition functions even on
gravitational tree level. Black holes become non-singular gravitational
instantons that solve the Euclidean Einstein equations. For this
transformation one needs to remove conical singularities that appear at
the event horizon (of non-extremal black hole solutions). This enforcing
of non-singularity of the solutions requires the imaginary time to be
periodic. Statistical mechanics tells us that a period in time is the
inverse of a temperature. Here holds $period = 1/T_{Unruh/Hawking} $ 
because of the relation between the periodicity of the Euclidean Green's
function and the role of the corresponding Green's function in
Lorentzian spacetime (one speaks of a thermal propagator). One can say
that any event horizon together with quantum field theory leads to
thermodynamics because quantum field theory in a Euclidean spacetime
with periodic time is equivalent to finite temperature quantum field
theory in Minkowski spacetime. 

In this method there appears a so called ``reference action'' subtraction
for the gravitational contribution on tree level. This subtraction is
needed in order to get finite results on shell. The justification for
this subtraction is that the action for a flat spacetime should vanish.
This reminds of the subtraction of the zero-point energy in quantum
electro dynamics. Therefore ~\cite{Belg96} suggests a gravitational
Casimir effect. The entropy of black holes could be due to this effect
and therefore it could be ascribed to vacuum fluctuations (zero modes)
at the event horizon and it would be natural that it comes with a
dependance on surfaces ( $ A_{+} $ ).   

$T_{H}$ one can obtain with the above outlined periodic in Euclidian time
$(x^0 \rightarrow i\tau)$ partition function method:
\begin{eqnarray}       
Z=\int D\phi_{(x)} e^{-S_{E(\phi)}}\\
\phi_{\tau + \beta} =\phi_{\tau}
\end{eqnarray}
where the period of $\tau$ gives the inverse of the temperature.
 For the Schwarzschild solution
for example we get a flat metric of a conical singularity with
\begin{eqnarray}
\frac{\beta}{2r_{+}} = 2\pi \, \Rightarrow T_{H} = (8 \pi M)^{-1} \label{TSSS} 
\end{eqnarray}
(compare with equation \ref{ab} ). 
For the charged black hole the Euclidian section leads to:               
\begin{eqnarray}
T_{H} = \frac{2 \sqrt{M^2 - q^2 }}{4 \pi (M + \sqrt{M^2 - q^2 })^2 }
\label{TRN}
\end{eqnarray}
which is $(8 \pi M)^{-1} $ for $M>>q$ but goes to zero for $M
\rightarrow q$ . The extremal black hole has zero Hawking temperature,
i.e. it is stable.          

As we will see shortly one can even define a chemical
potential associated with gauge charges and the usual relations of
thermodynamics are valid.
There are three ways to the entropy of a black hole:
\begin{itemize}
\item The fundamental one (based on the fundamental definition of
      entropy):
      Compute it directly from the saddle point approximation to the
      gravitational partition function (namely the generating functional
      analytically continued to the Euclidian spacetime as outlined above)
	~\cite{Gibb77}. This is a quite complicated way though.
\item Knowing $T_{H}$ and chemical potentials (i.e. the Coulomb
	potential at the event horizon ${\phi}_{+}$ for the conserved U(1) charge
      $q$ and angular velocity of the event horizon ${\Omega}_{+}$ for conserved
      angular momentum $J$) one may integrate the first law of
	thermodynamics 
	(compare with equation (\ref{firstlaw}))
	\begin{eqnarray}
      T_{H}\,dS=\,dM-{\phi}_{+}\,dq - {\Omega }_{+} \, dJ
	\end{eqnarray}
      to obtain $S$ ~\cite{Gibb77} ~\cite {R.Wald.84}. For the
	Schwarzschild solution for
	example one obtains the Bekenstein-Hawking formula (\ref{BHS})
	using equation (\ref{TSSS}): 
     \begin{eqnarray}
        T_{H}\,dS=\,dM \Rightarrow dS =(8 \pi M) \, dM 
        \Rightarrow S=4 \pi M^2 = \frac{4 \pi (2M)^2 }{4} =
	\frac{A_{+}}{4} \label{M}
     \end{eqnarray} 
	More precisely:
	\begin{eqnarray}
	dM = \frac{1}{2M} \, d(M^{2}) \\
	r_{+} = 2M \\
	A_{+} = 4 \pi r_{+}^2
	\end{eqnarray}
	lead to 
	\begin{eqnarray}
        dM = \frac{1}{8 \pi M} \, d(\frac{A_{+}}{4} ) = T_{H} \, d(\frac{A_{+}}{4} )
        \end{eqnarray}
	Assuming that the black hole exchanges mass (= energy) only trough heat
	$(dM=\delta Q) $ then $(\delta Q = T \, dS)$ holds for processes
	in thermal equilibrium. It follows
	\begin{eqnarray}
        dS = d(\frac{A_{+}}{4} )
        \end{eqnarray}
   	and with $ A_{+} =0  $ for $ M = 0 $ and the assumption that $S_{M=0} = 0 $
	follows 
	\begin{eqnarray}
        S = \frac{A_{+}}{4} \label{BHS}
        \end{eqnarray}
	The same holds for the charged black hole using equation
	\ref{TRN}:
	\begin{eqnarray}
	T_{H} \, dS = \, dM \, \Rightarrow dS = \frac{2 \pi  (M + \sqrt{M^2
	-q^2}
	 )^2}{ \sqrt{M^2 - q^2 }} \, \, dM \\
	= d[ \pi (M + \sqrt{M^2 - q^2})^2] = d[ \pi r_{+}^{2}]\\
	\Rightarrow S= \frac{A_{+}}{4}
	\end{eqnarray}
	which holds as well for the extremal black hole (there $ S = \pi (M^2
	+ a^2 )$)
	although $T_{H} = 0$ , as discussed above (Degeneracy of the ground state).

	This method can be hard if the
      expression of $T_{H}$ is complicated like it is for example
      in Brans Dicke gravity. 
\item Just take the Bekenstein-Hawking relation ~\cite{Bekenstein73}
	~\cite{Hawking76},
 	that is equation \ref{BHS}, which is 
      established in the context of \emph{Einstein's} general relativity via the
      first law of thermodynamics, see in ~\cite{R.Wald.84}. This is not
      valid in every gravity theory like for example Brans Dicke gravity.
\end{itemize}
  
Since Brans Dicke gravity theory ~\cite{Brans61} is not ruled out by
experiment and superstring theory can incorporate many forms of gravity and is about
compactifications (in Kaluza-Klein 5 dimensional theory the compactification of the
fifth dimension leads to the Brans Dicke scalar $g_{44}$) one should
speak more of entropy and not reduce it to the Bekenstein-Hawking one
as often done. This is to say that all is not easy and superstring theory is expected
to alter gravity theory of course, which might lead to Brans Dicke gravity or
something else and will most probably lead to the result that the
Bekenstein-Hawking entropy formula is only valid for large black holes.
We will see that the superstring results match the Bekenstein-Hawking
entropy only for configurations with many membranes and excitations. To
my knowledge, the first explicit statement of a ``correction'' to the
Bekenstein-Hawking temperature is due to ~\cite{Sen95}.  
In relation to our subject of black holes and thermodynamics it might interest the
reader that the non trivial Brans Dicke black hole solutions that were thought
impossible  ~\cite{Hawking72} and that have been reinstated in case Hawking's
weak energy condition fails 
~\cite{Campanelli93} are probably censored. At least the
semiclassical treatment censors, which means here that it leads to
 infinite $T_{Hawking}$ for them
~\cite{Kim97}.  Therefore Brans Dicke gravity theory is still viable without even leading to any
different black hole solutions.\\

The found entropy $S=\frac{A_{+}}{4} $ is very large indeed. With for
the Schwarzschild black hole $A=4
\pi r_{+}^{2} $ , $r_{+} = 2 M G $ and $m_{Planck} = G^{-1/2} $ ($m_{Planck} =
\sqrt{\frac{\hbar c}{G} } $ in SI units) follows:
\begin{eqnarray}
S=4 \pi (\frac{M}{m_{Planck}} )^{2} \label{Planck}
\end{eqnarray}
Using well known formulas for the energy and entropy of black body
radiation\\ (i.e. cavity radiation in a cavity of volume $V$) 
\begin{eqnarray}
E=4 \sigma VT^4 \\
S= \frac{16}{3} \sigma VT^{3}
\end{eqnarray}
,where $\sigma $ is the Stefan-Boltzmann constant, it becomes evident
that an astronomical black hole (say of a few solar masses) can never
fully evaporate because its entropy is so large that $V$ has to be
larger than the whole universe.\\

It should be noted that the Bekenstein-Hawking area law cannot always be
applied to extremal solutions. Some extremal black holes have zero
entropy but finite $A_{+}$ ~\cite{HawHor956}.\\
~\cite{Liberati97} suggest
that this is resolved by closer study of the topology of black hole
solutions --- or better their corresponding instantons in the Euclidean
metric. Proposed is the following entropy-area-law: 
\begin{eqnarray}
S=(\frac{\chi }{2} ) \frac{A_{+}}{4} \, \, \, \, \, 
; \chi = 0,2,4,\ldots 
\end{eqnarray}
, where $\chi $ is the Euler number of the manifold. For a $ d=4$ manifold holds
~\cite{Liberati97}:
\begin{eqnarray}
\chi = \sum _{n=0}^{4} (-1)^{n} B_{n}
\end{eqnarray}
where the $B_{n}$ are the nth Betti numbers. They show that $\chi = 0$
for extremal black holes and $\chi = 2$ for the non-extremal ones in
classical general relativity.
Later on, in the string theory part, we will find black holes with
$S\neq 0$ due to a microstate description but $A_{+} = 0$. For them
the Bekenstein-Hawking entropy will only be valid if thought to be due
to the ``stretched horizon'' about one string length further out than the event horizon. 
Extremal but non-zero entropy black holes have been found by
~\cite{Sen95} and ~\cite{S+V96}. This discussion on extremal black holes
(whether $S=0 \neq A_{+} $ or $A_{+} =0 \neq S $ ) is far from resolved but
~\cite{Ghosh9697} have shown that the superstring theory results will be
confirmed if one uses a summation over topologies and applies the
extremalety condition late (after quantisation).

The semi-classical approach to black hole thermodynamics is not a closed case   
and new results from superstring theory sparked new interest into
deriving thermodynamics via path integrals in non-trivial spacetime
topologies; eg.: ~\cite{Ortiz97} who argue that black hole topologies
are multipli connected:
\begin{quotation}
``\ldots , a particle may tunnel across the
horizon, and the topology of the whole configuration space is then
physically relevant. The radiation of a black hole is thermal because,
from the point of view of a distant inertial observer, there is a
denumerable infinite number of ways for a particle to tunnel through the
horizon \ldots ''
\end{quotation}

\section{Information Loss Paradox and Black Hole Complementarity}
The thermal spectrum of a star is the result of an averaging over its
microstates. Given a star in a pure state the radiation is actually
dependent on the microstates and correlated. A detailed analysis of the
radiation could (in principle) be used to determine the initial state.  
Say that we consider the collapse of a star in a pure state that is describing its
many degrees of freedom resulting in a black hole (say a Schwarzschild one). This
black hole still is a pure state
and the laws of quantum mechanics tell us that its phase relations evolve unitarily.
The pure state will always remain a pure state. The Hawking radiation on
the other hand 
will lead to a complete evaporation of the Schwarzschild black hole and the result of
this process is purely thermal radiation that cannot depend on the
initial state because the Hawking radiation depends only
on the outside geometry of the black hole but the information carrying matter is
inside the black hole in the framework of classical general relativity.
This means that the
phase relations are lost completely and in principle; the final state is
not pure but mixed. The information is
not encoded in correlations among the particles of the Hawking radiation
because this would violate principles of field theory (locality and
causality especially when the information is thought to be in the middle
of the black hole at the singularity) and basically goes against the whole notion
of thermal radiation. The violation of unitarity in time evolution that 
we encounter here is called the information loss paradox because we
lose all information (except parameters like the mass $M$) once the
information carrying matter falls through the event horizon. Hawking defends his
point of view ~\cite{Hawking7576} that the evolution of black holes is not unitary and that
quantum mechanics has to be altered. Hawking proposed the
``superscattering operator'' $ \$ $ that acts on the density matrices
$\rho $ in such a way that the state functions are not evolving due
to an unitary S-matrix $S$:
\begin{eqnarray}
\rho_{f} = \$ \rho_{i} \, \, but \,\,  |\Psi _{f}> \neq  S |\Psi _{i}>
\end{eqnarray} 

Other aspects of the information paradox are the problem of where the
great amount of entropy of the black hole is stored after the Hawking
evaporation and the violation of conservation laws. The latter comes
from the fact that the black hole has lost all information about how
many baryons and leptons for example have fallen into it. The former is
basically saying that our universe is too small to allow for evaporation
of black holes. Practically the temperature would never drop low enough
to let the black holes evaporate fully. ~\cite{Mash97} claims that both
aspects are due to the assumption that the black hole loses energy 
through heat only. The requirement of conservation laws even for the
black hole via 
\begin{eqnarray}
(dE = \delta Q) \rightarrow (dE \, = \, \delta Q \, + \, 
\sum_{a} \mu _{a} N_{a} )  
\end{eqnarray}
with chemical potentials ( $\mu _{a} $ ) for the baryons etc. leads to
the possibility of entropy loss that is not heat related and to 
less entropy for the black hole with a correction to the
Bekenstein-Hawking relation ~\cite{Mash97}:
\begin{eqnarray}
dS= \frac{dA_{+}}{4} - \frac{1}{T} \sum_{a} \mu _{a} \, dN_{a}                  
\end{eqnarray} 

It should be noted that the information paradox is a purely ``inner
theoretical'' one. For the observer outside matter never
actually falls through the event horizon because of time dilation.
Shortly before the collapse, the information from the region inside the
collapsing region has no chance of
escaping anymore. Therefore, the event horizon comes always out of the
middle and even things in the centre are not seen to ever fall through.

In \cite{SUSK931}, where the notion of the ``stretched horizon''is 
introduced, one can read most clearly about the complementarity between
observations of distant observers and observers who fall through the   
event horizon. In a nutshell: One either describes the physics from    
outside the black hole or from the inside but because no information can
be exchanged between these two observation posts it is nonsense to
require a description valid for both. Such a description would have no
operational meaning.  I like this complementarity suggestion because it
goes hand in hand with ``the world'' and its physics being
whatever is self-consistently possible for a consciousness as such. This
point of view does not exclude but does not require a description valid 
for all possible observers at once either. In the background of a black 
hole, quantum field theory as we know it today leads to states
describing the inside and the outside at once
and a quantum field theory of gravity
certainly would do so.\\
An observer falling freely through the event
horizon will observe no shell or membrane at all if the equivalence
principle does hold there. But for our purposes of understanding black
hole thermodynamics one might like to adopt the complementary side and
view the black hole from the outside seeing the event horizon as a physical
membrane  ~\cite{SUSK931} page 3745:
\begin{quotation}
`` The membrane is very real to an outside observer. For example, if such
an observer is suspended just above the stretched horizon, he or she
will observe an intense flux of energetic radiation apparently emanating
from the membrane. If provided with an electrical multimeter, our
observer will discover that the membrane has a surface resistivity of
377 ohms. If disturbed, the stretched horizon will respond like a
viscous fluid, albeit with negative bulk viscosity. And finally, the
observed entropy of the massive black hole is proportional to the area
of the stretched horizon.''
\end{quotation}

Anyway, one is free to suspect that the theory of general gravity does
not break down near the singularity it predicts but near the event 
horizons. That would mean that we cannot fall freely through the horizon. 
 
With the recent advances in superstring theory one is able to address
the information paradox. There are claims that the problem is now solved
 ~\cite{Amati97}, that the
evolution is unitary and that the Hawking radiation is only thermal
when computed in the classical limit. Basically, the (any) thermal radiation only looks
thermal when quantum mechanics is fully applicable because  thermodynamics
is about observation (measurement) and the microscopic descriptions one
seeks are to extract the probability amplitudes that an observation will destroy. 

\newpage
\part{String Theory of Black Holes}
\section{Introduction to the New Approach}   
\subsection{BPS States and Dualities}
This is not the place to introduce supersymmetry but we will at least need to
know a little about BPS States later on. BPS means satisfying the 
Bogomol'nyi-Prasad-Sommerfeld bound that (given appropriate normalisation) the
charges  equal the self-coupling (mass). More precisely, the mass is
bound from below due to the supersymmetry algebra having conserved
charges that are not momentum or supercharges. A BPS-state has a mass
equal to this lowest bound (it ``saturates'' the bound). Therefore, BPS
states are in reduced multiplets of the supersymmetry algebra (short representations)
and the supersymmetry
protects their charges (like mass) from changes due to changes of
coupling constants (from weak to strong for example). This has its
parallel in the photon/Z-particle for which $m \geq 0$ and that is in a
short representation of the Pointcar\'{e}  group if $m=0$ because the longitudinal polarisation is 
missing.  Recall that $q=M$ was the condition for an extremal black hole
as well.\\

T-duality means simplified that compactification of one theory 
for  example with radius $R$ gives the same spectrum of states as the
$ \frac{ \alpha '}{R}$-compactification on the dual lattice of 
the thereby T-dual theory (see equation (\ref{Mdepnm})) and 
the same interactions if $g \rightarrow g \frac{ \sqrt{\alpha '}}{R}$.
The unit cell of the lattice has volume $V$ and the dual lattice $1/V$
(for compactification of one dimension only $R \leftrightarrow  
1/R$ (simplified)). Therefore, T-duality translates into ($ \psi
\leftrightarrow - \psi $) if we express the volume $V$ ($R$ in one
dimension)
as the expectation of a scalar field $\psi $ :
\begin{eqnarray}
<e^{\psi }> = V
\end{eqnarray}
Note that the purely mathematical duality relations $(V \times (1/V) =
1)$ are given. We have to translate this into less beautiful relations
between coordinate space and momentum space for which winding and
quantisation of momentum make sense. With volume $= \prod_{i}(2 \pi R_{i})$ :
\begin{eqnarray}   
((2 \pi )^{d} V) \times (\frac{1}{V} \hbar ^{d}) = h^{d}
\end{eqnarray}
We need $g \rightarrow g \frac{ \sqrt{\alpha '}}{R}$ for the
interactions to be invariant under T-duality because we will see that:
\begin{eqnarray}
G_{N}^{(10)} \propto g^2 (\alpha ')^4 \label{Gg} 
\end{eqnarray}
With 
\begin{eqnarray}
G_{N}^{(d)} = \frac{G_{N}^{(10)}}{V_{T^{(10-d)}}} \label{Gd} 
\end{eqnarray}
where $V$ is the volume of the torus of compactification one needs 
$g \rightarrow g \frac{ \sqrt{\alpha '}}{R}$ in order to leave 
$G_{N}^{(d)} $ invariant. For example:   
\begin{eqnarray}
G_{N}^{(9)}
 = \frac{8 \pi ^{6} g^2 (\alpha ')^4 }{2 \pi R} \rightarrow 
\frac{8 \pi ^{6} (g^2 \alpha ' /R^2 ) (\alpha ')^4 }{2 \pi (\alpha ' /R)}
= \frac{8 \pi ^{6} g^2 (\alpha ')^4 }{2 \pi R} = G_{N}^{(9)}
\end{eqnarray}

S-Duality \cite{Sch93} is strong-weak coupling duality
like for example electric-magnetic duality
\begin{eqnarray}
g \leftrightarrow 1/g \, \, \,  and \, \, \, 
R \leftrightarrow \frac{R}{\sqrt{g}} \label{Sdual}
\end{eqnarray}
with the latter for all radii of compactifications that there happen to be.
It is very similar to T-duality (simplified $R \leftrightarrow
1/R$) especially since coupling constants can be radii of
compactifications (see equation (\ref{a})). In fact, the expectation value
of the dilaton field $\phi$ fixes the dimensionless coupling of
superstring theory
\begin{eqnarray}
<e^{\phi}>=g  
\end{eqnarray}
thus the coupling is a moduli just like the radii of compactifications are. 
S-duality translates as $ \phi \leftrightarrow - \phi $. Therefore it is
convenient not to use the string metric $G_{\mu \nu } $ that will appear
in the supergravity equations but the Einstein metric 
\begin{eqnarray}
g_{E} = e^{-\phi /2} G  \label{EM} \\
(g_{\mu \nu} = e^{-4\phi /(d-2)} G_{\mu \nu} )
\end{eqnarray}
that is then used to evaluate ADM masses and 
which is invariant under the S-duality transformation because of the
$g$-dependence of $G$. 
 
U-duality mixes S-  and T-dualities and translates as $( \psi
\leftrightarrow \pm \phi$). There are two consistent quantisations
possible for the left and right moving spinors on a closed string that
is parameterized by $0 \leq \sigma \leq 2 \pi $ :
\begin{eqnarray}
Ramond \, (R):          \Psi ^{\mu}_{\sigma + 2\pi } = + \Psi
^{\mu}_{\sigma } \\
\, \,  Neveu-Schwarz \, (NS):   \Psi ^{\mu}_{\sigma + 2\pi } = - \Psi
^{\mu}_{\sigma }
\end{eqnarray}
With an independent choice for the left and right movers there are two 
bosonic sectors (R,R and NS,NS) and two fermionic ones (R,NS and NS,R).
One remarkable fact about U-duality is that it unifies R,R and NS,NS  
sectors ~\cite{Hull95} ~\cite{Hull951}. This shows that the difference
between periodic and anti periodic boundary conditions for fermions  
is related to the shortcommings of perturbative string theory
and vanishes at a certain level in the non-perturbative theories.
When we discuss D-branes we will see more closely how R,R charged
D-branes are turned into NS,NS branes. They appear in the same U-duality
multiplet ~\cite{Schwarz950}.

Recall 
that BPS states are in short representations. This means that they have 
properties that do not change due to renormalisation ~\cite{Witten782}  
~\cite{Kallosh92} (they do not get quantum corrected, i.e. do not depend
on the coupling strength) and stay constant even when other moduli than
the coupling strength are changed. That BPS states are coupling
independent brings with it the appearance of them in any theory that is
connected via dualities. Thus, BPS states have been vital in testing  
non-perturbative dualities ~\cite{Hull95} ~\cite{Witten952}. For black
hole calculations it will be important that the validity of results can
be extended
into strong coupling ~\cite{Sen95} .  Charges will be counted in order  
to give the degeneracy due to microstates and the mass gives the entropy
due to the area of the event horizon. In order to compare then in a
meaningful way in another coupling regime one requires that the
mass-charge relation stays the same. That will be so if the states stay 
in their short representations, i.e. are BPS. For p-dimensional branes  
the BPS inequality is one between the tension $T_{p} = mass/volume$ and
the charges. For BPS states there is a projection operator
($\epsilon ^2 = \epsilon $ ) such that
\begin{eqnarray}
\epsilon Q |BPS> = 0
\end{eqnarray}
where Q is the supersymmetry charge. This is the very meaning of the  
statement that the presence of a BPS state preserves
$1/a$ with $a=2,4,8$ of the supersymmetries.

\subsection{Superstring Theories and M-Theory}
The great interest into  string theories comes partly from
the fact that they give rise to a lot of symmetries 
and related non-perturbative methods. The latter is due to dualities and
solitons. Introductions to string dualities are: \cite{Sch951}
\cite{Pol961} \cite{Sch961}.
String theories, just by being theories not about matter
points but about the simplest possible extended objects, generate
gravity consistent with quantum mechanics and they are the only theories yet we know of
that do so. Moreover, starting with just a few postulates, superstring theories provide
easily what GUTs always wanted to give but GUTs always do so in a more
artificial manner leaving a lot of free parameters like the choice of
the gauge group. Superstring theories generate gauge groups large enough to incorporate
all gauge groups of the standard model leaving only very few degrees of
freedom that are well understood geometrically: The string scale length,
and compactification parameters; the latter hopefully being fixed by
selfconsistency eventually. Other assets of string theories are that we
find finite theories, axions, bounds on the number of lepton
generations, light Higgs bosons addressing the hierarchy problem 
and maybe the most important asset is that
string theory needs supersymmetry and will most probably be testable via its
predictions of ``low'' energy supersymmetric effects. The first
superstring revolution (1984) led to a most welcome bound on the
dimensionality of spacetime by showing that there are five consistent
superstring theories and all are in 9+1 dimensions. This is ``most
welcome'' because such theories might explain why we perceive a 3+1
dimensional spacetime. 

However, the euphoria among some particle physicists who put superstring theory close
to the endpoint of all fundamental physics is not quite justified.
Problematic is that superstring theories still use the concept of something very
unphysical --- the spacetime point. Nothing is able to resolve a point
and a fundamental theory should have some characteristics of for example
twistor theory where one gets rid of spacetime points after quantisation
is done in twistor space. Twistors are generalisations of spinors. With
fundamental spinors (only up or down) one can build up an $O(3)$
symmetry. In order to obtain Poincar\'{e} symmetry one needs to somehow
add the aspect of momentum to the spinors. The result is called ``twistors''.
Indeed, twistorial methods in superstring theories and
supergravity theories were advocated by Witten ~\cite{Witten78}.
 A singularity often just indicates a point
where a theory is breaking down.  Similar to the interpretation that
quantum mechanically smeared out 
electrons do not ``see'' the singularity of the electric field of a
nucleus (and therefore do not spiral inside) there is an interpretation
 saying that the finite length of the
string makes it blind to the singularities of gravitational fields and
thus general relativity and quantum mechanics are reconciled.  This
argumentation still accepts
spacetime singularities of background fields and the extended objects
have zero thickness and moreover, they can be wrapped up along
compactified dimensions (isospaces) so that in \emph{the} spacetime
(the uncompactified one) there is nothing left but a point. Superstring theory leads
in the right direction though since there are proposals that the
curvature singularity inside a black hole is indeed string theoretically just
another coordinate singularity through which matter passes smoothly
(because there are dualities that interchange event horizon and singularity)
\cite{Giveon91} \cite{Dijk92}. Others suggest that the
singularity is not there at all and that the black hole is homogeneous
inside \cite{Hotta97}.

Other downsides that we might be able to overcome shortly are the many
vacua of superstring theories: some come as discrete choices, some even
as continuous ones.
Problems with the weak coupling limit led to the now accepted view that
so called M-theory is the more fundamental one with the superstring theories being
certain limits of it. Dualities allow us to formulate strong coupling
problems as weak coupling ones. Only at weak coupling one can perform
real calculations and these lead often to wrong predictions like
unstable vacua. There are even general arguments against the validity of
any superstring theoretical weak coupling analysis.  

Instead of weakly coupled strings one now favours \cite{TOWN97} 11 dimensional
M-Theory\\
\cite{Sch952} \cite{Duff96} \cite{Town96} 
which features membranes of higher dimensionality than strings
but which has a non-perturbative only strings-formulation (only open strings and
D-0-branes (see later)) called (M)atrix theory\\ ~\cite{Banks96} which is
incomplete in case many dimensions are compactified. Although it is a
matter of dispute, the name ``M-theory'' seems to come
from the now obsolete view that it is a theory of two dimensional
branes (Membranes) only. The strings would appear after compactification of the
eleventh dimension.
The low energy limit of M-theory is 11 dimensional supergravity.
The 5 consistent  (anomaly free) superstring theories
represent different corners (vacua) in a large phase diagram
that is supposed to describe M-theory. That all five are related by
dualities (fundamental states in one are solitons in a dual description)
one has known before (second superstring revolution 1994). Now one can view the
superstring theories as certain 
compactifications of the more general 11 dimensional theory
\cite{Sch962} \cite{Sch963}. The latter or maybe an even more general
theory [Since IIB superstring theory is not accessable via
a straight compactification of 11 dimensions --- that is only possible
from 12 dimensions (F-theory ~\cite{Vafa96})] should give rise to a big so called
moduli space since the dynamical fields whose expectation values are the
parameters characterising different superstring theories (e.g.:radii and the shape of
compactifications) are called moduli. 
How fundamental is M-theory ? ~\cite{Witten97} has shown that the 5
dimensional membrane of M-theory can be used to model supersymmetric
gausge theories in four dimensions precisely. In fact, the
series [IIA/B in 10, M in 11, F in 12 dimensions] is mirrored quite precisely by so
called ``little superstring theories'' called [a/b in 6, m in 7, f in 8
dimensions]. This points to the possibility of a whole tower of theories
([6,7,8 with gauge summetry but without gravitation], [10,11,12 with
general covariance], [14,15, 16 with ???], \ldots ) in which M-theory might
be due only to one particular 11 dimensional membrane \cite{Losev97} .

However, all we know for certain about M-theory is its weak coupling
limit, the 11 dimensional supergravity. The latter is not renormalisable
(has UV-divergences) but
M-theory hopefully is by being the embracing structure visible only at 
strong (infinite) coupling. Compactifying the 11th dimension onto a
circle $S^{1}$ of radius $R_{11}$ gives ten dimensional
type IIA superstring theory which (at weak coupling) has BPS solitons with mass 
\begin{eqnarray}
M^2 = \frac{n^2}{g^2}\, ; n \, \, integer \label{mdepgsq}
\end{eqnarray}
These IIA BPS soliton states can now (from M-theory) be understood as
winding states, here with
\begin{eqnarray}
R_{11} = g^{3/2} \label{a}
\end{eqnarray}
since a $S^1$
compactification leads to formulae like 
\begin{eqnarray}
M^2 \propto ( \frac{n^2}{R^2} + \frac{m^2 R^2}{(\alpha ')^2})   \label{Mdepnm}
\end{eqnarray}
(simplified) where n is due to the momentum quantisation
along $S^1$, $p=n/R$, and m is due to the winding. That these winding
states can be BPS (n \emph{or} m equal to zero) comes from the fact that
the momenta along the eleventh dimension appear in the supersymmetry
algebra and are therefore turned into central charges (winding and
momentum along $x^{10}$) after compactification. The components of
the 11 dimensional metric lead to: $g_{\mu ,10} = A_{\mu}$ and
$g_{10,10} =R_{11}^{2}$. Thus, as mentioned before, the radius is the
expectation of a moduli and is dynamical although it can have any value,
i.e. there is no potential for it.  Moduli have in lowest order
approximation no potential and if there is enough supersymmetry there will be
perturbatively none at all (of course, non-perturbatively there has to
be some mechanism that fixes them). 

The expectation value of the dilaton field $\phi$ fixes the
dimensionless coupling of superstring theory
\begin{eqnarray}
<e^{\phi}>=g  
\end{eqnarray}
thus the coupling is a moduli just like the radii are and from the
M-theoretical point of view the D=10 dilaton is just a component of the
D=11 metric.

$S^1 / Z_{2}$ compactification on an interval of the 11th dimension 
( i.e. the 11 dimensional manifold has two boundares separated by
$R_{11}$, no
multiple windings) leads to heterotic $E8 \times E8$ superstring theory
\cite{Horava96}. After this compactification fields of the 11
dimensional supergravity (e.g.: graviton) propagate through all 11
dimensions but the gauginos and gauge fields are restricted to the
boundaries. This is an important fact because it makes the gravitational
coupling grow faster than the three couplings of the standard model and
all couplings meet at the same \emph{unification} scale if the length of
the interval is chosen appropriately.\\ 

With these duality connections one can relate all 5 consistent superstring
theories  to
M-theory because type IIB and heterotic SO(32) are related to IIA and
heterotic $E8 \times E8$ respectively via T-dualities and type
I SO(32) is related to heterotic SO(32) via S-duality (IIB is
S-self-dual).

According to \cite{Sen96} the type I / heterotic duality (based on a
duality between strings and five dimensional membranes) is enough to
generate all other relations, including relations of M and F theory. 
The equivalence of type I (a theory of open unoriented strings) and the
other four superstring theories (all of oriented closed strings) tells
us  that the distinction between open and closed string theories is not
that important anymore. In general, duality relations allow us to be
rather free in our choice of objects we use for models. Because of this
and because of the closeness of theories with objects of higher dimensionality
and superstring theory, the subject of black hole entropy
is said to be addressed by string theory although the vital tool will be
D-branes and M-theory eventually.

\subsection{D-Branes}
String theory becomes more and more a theory about p-dimensional
membranes. The usefulness of so called D-branes comes from them being
massive BPS states if the string theory is supersymmetric.

T-duality leads to certain new boundary  conditions
for strings and therefore naturally to D-branes ~\cite{Dai89}. Since
then the subject of dualites has become much clearer since D-branes give
a detailed dynamical description with the D-brane being in the duality
multiplets together with fundamental strings and solitons of field
theory. For the
early work in the context of bosonic string theory see \cite{Dai89}
\cite{Leigh89} \cite{Horava89}. To be recommended as a 
review on D-branes is \cite{Pol962} and a very recent review is
~\cite{Thorlacius97} .
Stringtheory (M-theory)  has D-branes (Dirichlet p-dimensional membranes) ---
domain walls (solitons) of dimension 0 to 9 (D-0-brane is a point,
D-1-brane a string and so on.) necessarily ~\cite{Polch94}. ``D'' stands
for Dirichlet boundary conditions (see equation (\ref{Dirichlet})) 
of open strings. In fact, D-branes are defined via the
boundary conditions of strings ending on the brane and this is understandable as soon as
one realizes that the open strings are the excitations of the D-brane
which in turn is expected to be non-rigid because we are dealing with a
theory that incorporates gravity. An excited D-brane is described by a
gas of strings on the brane. For example: Massless bosonic open strings
have one Lorentz index and if the dimension of this index is
perpendicular to the D-p-brane then it is an oscillation of the brane.
If the index is parallel (inside the brane) then this corresponds to gauge
fields on the brane. The defining open strings end on the
D-brane and the string end points can move freely inside the brane but
satisfy Dirichlet boundary conditions in directions transverse to the
thereby defined brane. Recall that the variation of the free
string action with respect to the auxiliary field $h_{\alpha \beta}$
(the world sheet metric) and covariant gauging after that gives the open
string boundary condition as
\begin{eqnarray}
\partial_{\sigma}X^{\mu}=0
\end{eqnarray}
at the string ends. This is the well known Neumann boundary condition
saying that an open vibrating string has displacement nodes at $\lambda
/4$ from the string ends. For the 10 dimensional type II open strings
that are introduced into the theory of closed strings via the D-brane
these conditions hold for say $\mu = 0, \ldots ,p$.
The Dirichlet boundary condition 
\begin{eqnarray}
X^{\mu}=a^{\mu}\, constant \label{Dirichlet}
\end{eqnarray}
describes the location of the D-p-brane for say $\mu = p+1, \ldots ,9$.
In such a way open strings are consistently introduced if p is even for
IIA and odd for IIB superstring theory. This even/odd dichotomy arises
from the boundary conditions that open strings/branes require. Closed
strings do not relate left and right moving spinors but for open strings
the fact that IIA/B has spinors of opposite/same chirality will lead to
restrictions on p that we find from a different point of view again
later on. That the left and right moving spinors become reflected and
mixed at the boundaries of the D-brane leads to them breaking half of
the supersymmetries. That is exactly the condition for BPS states and
thus D-branes are BPS. That 1/2 supersymmetries survive is similar 
to the fact that open string theory is type I because the boundary
conditions at the open ends mix right and left moving
spinors.\\

The D-branes are translation and boost invariant along dimensions
parallel to the brane.
This relativistic behaviour comes from the preservation of some of the
supersymmetries and demands that the longitudinal momentum of
the D-brane is modelled by the excitations of the branes (open attached
strings) that move along the brane with light velocity --- the latter  
because the tension of the D-brane is $T_{P} = mass/volume $ just like
it has always been for the fundamental strings.

D-p-branes couple to  $(p+1)$-dimensional antisymmetric tensor potentials
with gauge invariance $A_{p+1} \rightarrow A_{p+1} + \, d\phi_{p}$  with
$d\phi _{p} = \delta A_{p+1}$ the gauge transformation and
here Abelian field strength $F_{p+2} = \, dA_{p+1}$ or without
\begin{eqnarray}
A_{p+1} = A_{\mu_{1},\ldots , \mu_{p+1}} \, dx^{\mu _{1}} \, \wedge
\ldots \wedge   \, dx^{\mu _{p+1}} 
\end{eqnarray}
differential forms:
\begin{eqnarray}
A'_{\mu_{1},\ldots , \mu_{p+1}} = A_{\mu_{1},\ldots , \mu_{p+1}} + (p+1)
\partial _{[\mu_{1}} \phi _{\mu_{2}, \ldots ,\mu_{p+1}]}\\
F_{\mu_{1},\ldots , \mu_{p+2}} =  (p+2) \partial _{[\mu_{1}} A
_{\mu_{2},
 \ldots ,\mu_{p+2}]}
\end{eqnarray}
This holds for p-pranes in general (eg: the 0-brane $e^{-}$ couples to
$A_{\mu}$ with $F_{\mu \nu} = 2 \partial_{[\mu}A_{\nu]}$).
D-branes couple to these tensor gauge fields with the coupling constant
being an integer unit, the Ramond charge. Due to supersymmetry the
relationship $M=Q/g$ holds
precisely in the quantised theory. One says that they
interpolate between fundamental (NS) strings (Tension $T \propto g^0$) and (NS)
solitons ($T \propto g^{-2}$). There are two ways of calculating the masses of
D-branes. One way uses dualities rather intensively. Starting with a
fundamental string its minimal mass is given by its momentum
quantisation and/or winding along a compact dimension (equation
(\ref{Mdepnm})). SL(2,Z) symmetry
in low energy IIB (U-duality) relates this fundamental string to a
series of soliton solutions (all in the ``SL(2,Z)-multiplet''
\cite{Schwarz950} ) in the following way: Using
S-duality the fundamental string turns into a D-string. Remember that
the Einstein metric is invariant under S-duality and the metric is used
for the determination of masses. Thus, the mass stays the same appart
from factors like  $g$ for example which is due to the
difference between Einstein and string
metrics ($ \Rightarrow $  mass of D-string $\propto 1/g $). T-duality
perpendicular to a D-p-brane creates a D-(p+1)-brane. Therfore, from the
minimum mass for the D-string one can deduce the minimum masses for
D-branes of higher dimensionality. Now one can go back to NS states via
a further S-duality. 

The other way is computing D-brane masses via virtual closed string
diagrams of the mutual gravitational attraction between D-branes. When
dualities were still being discovered it was perceived as one of the
miracles of string theory that these two ways lead to the same mass
\cite{P95}. The results for the brane tensions are as follows:
\begin{eqnarray}
T \propto \frac{1}{(\alpha ')			} \label{Tfund} \\
T \propto \frac{1}{(\alpha ')^{(p+1)/2} g	} \label{tensions} \\
T \propto \frac{1}{(\alpha ')^{3}       g^2	} \label{Ts5} 
\end{eqnarray}
for the NS-fundamental string, the D-branes (RR-solitons) and the
NS-solitonic 5-brane respectively.

All this makes the D-brane models rather simple. The relation between 
D-branes and some T-dualities leads to a very useful ``D-brane technology''
because T-duality is an exact symmetry and one understands it
perturbatively.  S-duality is used less often in this context.  
D-branes give the right model for many objects that are required by the
dualities. So called D-instantons \cite{Corrigan75} \cite{Green76}  
\cite{Cohen86/87} have Dirichlet boundary conditions for all $\mu$ 
including time. D-instantons can probe
sub-string and sub-Planck scales. Thus the D-branes point towards degrees
of freedom on an otherwise as meaninglessly short discarded length scale.
On how D-brane-D-brane scattering at D-instantons probe shorter distances
than the Planck or string length see \cite{Shenker95}. 
Thus, the Planck scale is probably not the
smallest length but the one where quantum mechanics and general
relativity (these are the theories that
give the constants contributing to the Planck length) have to be
considered together as one theory giving a new geometry (non-commutative
geometry) below this length. There has been quite a lot of work on
D-brane-D-brane scattering and the emergence of very short scales:
\cite{Kleb95} \cite{Bach95} \cite{Barb96}
\cite{Douglas961} \cite{Douglas962}. 

Why do D-brane considerations lead to non-commuting geometry ?
D-branes are massive BPS-states. Therefore they gravitate or in general
interchange closed strings  ---  for
example gravitons of course. When the distance between the membranes is
smaller than the string length scale the description via closed strings
is not convenient anymore because it involves the exchange of all the
massive modes of the strings of which there are infinitely many.
Basically, if the branes are so close that the average open string
does not fit in between them the description in terms of open string
exchange will be unphysical. Physical is that the open strings squeezed
between two branes that approach each other will touch both branes and
break up. Therefore, the short distance interactions are best described
by  taking
virtual open strings between the D-branes to model the propagation of
a closed string. These open strings are gauge fields transverse to the
branes. $N$ D-branes close to one another (for an exact statement read: ``on
top of each other, identical'') give a $U(N)$ Yang-Mills interaction
because the connecting open strings (massless if D-branes have zero
separation) have ``Chan-Paton'' indices $(i,f)$ with $i,f \in \{ 1, \ldots
N \} $ indicating which D-brane a given string connects and where the
oriented string starts and ends. A single brane has a $U(1)$ gauge field
in it ($i=f $ for all $(i,f)$).
The distance of the branes is now expressed in terms of the open strings
connecting them (The coordinates of these strings are spacetime
coordinates as well.). This gives matrices that not necessarily
commute. Thus, below the string length scale we find the quantum gravity
being a non-Abelian gauge field theory. At
large distances the geometry becomes Abelian again. The branes
disconnect, the group $U(N)$ of the Chan-Paton factors $(i,f)$ becomes $U(1)^{N}$
and the location of the branes is still
described by matrices but they are now diagonal ones and do commute
~\cite{Witten95} . The issue of scales (string, Planck and GUT) 
is quite intensively debated. There are many 
suggestions about what one encounters near and below these scales 
and whether and which scales are identical. There could wait an entirely
new geometry, smaller elementary units or a phase transition. For hints  
pointing to the latter see for example ~\cite{Dick97})

In (M)atrix theory membranes 
(including strings) are viewed as composites of D-0-branes which
carry a single unit of RR-charge (Ramond-Ramond). The membrane is a collective
excitation of the D-0-branes. The strings that connect very close
D-0-branes result in a ``coordinate space'' in which the 9 spatial
coordinates of $N$ D-0-branes become 9 $N \times N$-matrices
~\cite{Banks96}. The large $N$ limit corresponds to the large momentum
of the light cone frame. ~\cite{Banks96} conjectured: 
\begin{quotation}
``The calculation of any
physical quantity in M-theory can be reduced to a calculation in matrix
quantum mechanics followed by an extrapolation to large $N$.''
\end{quotation}
Therefore, it would be a non-perturbative form of quantum-gravity.\\

The M-theory point of view opens up new possibilities for relations
between strings and branes --- just like $g_{\mu \nu }$ and the three form
$A_{ \mu \nu \rho }$ in 11 dimensions lead after compactification to 10
dimensions to $g_{\mu \nu }$, $A_{\mu}$, $\phi = g_{10,10}$ and $A_{\mu
\nu \rho}$, $B_{\mu \nu}$ etc. respectively in IIA superstring theory.
We can now have for instance 2-branes in 11 dimensions wrapping up to
give a string of type II in 10 dimensions, or a D-2-brane of IIA theory
after dimensional reduction. 

The D-branes will be wrapped up in compactified dimensions in order to
give charged point particles for the observer in the uncompactified
spacetime dimensions. Calculations will show that these
charged localized particles have an event horizon that is inside
the string length scale so that they are not jet black holes for the
open strings attached to them or when interacting with other objects
 (D-brane scattering). Therefore one has to superimpose many branes and 
strings in order to obtain a black hole -- that leads to strong coupling.

In order to model a near extremal black hole and its
Hawking radiation one excites the D-branes of an extremal black hole.
This is basically nothing else but D-brane scattering at an extremal black
hole. The extremal black hole will absorb, become excited, that is a near
extremal one, and will then scatter back which means that the near
extremal black hole radiates Hawking radiation until it is an extremal
black hole again. A review on D-branes being probes of black holes is
~\cite{Malda97} .

\newpage
\subsection{Supergravity}
Before D-brane technology the most important instrument for
investigation of the non-perturbative aspects of superstring theory
was supergravity. Even
now, although there are the successes of (M)atrix theory, the low energy
limit of M-theory is \emph{the} other handle that we can be certain of allows 
to study M-theory. This limit is 11 dimensional supergravity. 

Supergravity contains scalar fields, U(1) gauge fields and fermionic fields
that are not the subject of classical general relativity but otherwise 
the theories are very
similar. $N=8$ supergravity has the exactly same Reissner-Nordstr\"{o}m solution
where there is only one gauge field excited. These supergravities can be
introduced via the introduction of supersymmetry into stringtheory. For example, 
$N=8$, $d=4$ supergravity is the low energy limit
of the type II superstring theory compactified on
the torus $T^{6}$. The strings and p-branes of the superstring theory can be black,
i.e. they are inside their own event horizons, and indeed if these black branes are charged
and wrapped up around compact dimensions than they will be just like a charged black
hole in the context of general relativity where in \emph{the}
uncompactified spacetime one would observe a spherical symmetric
(or axis symmetric in case of rotation) event horizon and suspect a point
(or ring) singularity within it.  

The important point in the context of black holes is that in
supergravity there are the BPS-states and they are stable states.
Moreover, many\footnote{Actually, the most extremal black holes are not BPS, for
example extremal $a \neq 0 \neq q $ solutions or see in ~\cite{Kall92}.}
extremal and therefore stable ($T_{H}=0$) black holes are
BPS as we discovered before (recall the charge-equal-mass condition).
One can compute variables like the entropy of an assembly of
D-branes (which are BPS as well) for weak coupling and then these
results are valid for any coupling strength --- even for the regime
where the D-brane assembly is an extremal black hole. Hence, the gravity theory
being ``super'' was vital for the calculation of extremal and near
extremal black hole entropy not only because the string theory should be
a superstring theory in order to be free from tachyons for example. We find
D-branes as well in bosonic string theory.
The near extremal calculations are dependent on the extremal ones in a
very close sense as we discussed before (D-brane scattering).

In practise, obtaining supergravities from superstring theories means 
expanding the closed string $\beta $ -functions and then keeping only the
lowest order in $\alpha ' $ terms (Regge slope). The supergravities have
UV-divergences and are not renormalisable but are much easier to handle
(therefore the name ``effective'' theory). Many properties of the
supergravities that we obtain by investigation of the low energy
effective action of a superstring theory follow straight from the
properties of the latter. From uncompactified IIA/B superstring theory
we obtain $N=2$ supergravities IIA/B with different/same chiralities of
the supersymmetry generators. For both the NSNS sector fields are the
same. They are the string metric $G^{\mu \nu }$ , the two form $B^{\mu \nu }$ with
field strength
$ H= \, dB  $ or ($H_{\mu \nu \rho} = 3 \partial _{[ \mu } B_{ \nu \rho ] }$)
and the dilaton $\phi $ and they lead to the same terms in the action that follows
as (compare with equation (\ref{action})):
\begin{eqnarray} 
S_{II} = ( \frac{1}{16 \pi G^{(10)}_{N}}) \int \, d^{10} x \, \sqrt{-G} \,
[ e^{-2 \phi } (R + 4(\bigtriangledown  \phi )^{2} - \frac{1}{3} H^2 ) -
\alpha ' \hat{H} ^2 + \ldots \, ] 
\end{eqnarray}
where ``$-\alpha ' \hat{H} ^2 +$ \ldots '' is different 
in IIA and IIB due to different RR fields
and all the fermionic fields. $G^{(10)}_{N}$ is the ten dimensional
gravitation (Newton) constant, $ \hat{H} = \, dA$ represents a RR field strength
and R is the scalar curvature again (not a radius of a
compactification). The differences between the RR
sectors of type IIA and IIB are as follows:

\emph{In IIA supergravity \cite{Campbell84} :} A is a one form 
($ H_{\mu \nu } = 2 \partial _{[ \mu } A_{ \nu ] }$ )
and there is $ F' = 2 \, A \wedge H \, + \, dC $ where C is a three form
($ F'_{\mu \nu \rho \sigma } = 8 A _{[ \mu } H_{ \nu \rho \sigma ] } + 
4 \partial _{[ \mu } C_{ \nu \rho \sigma ] } $). 

\emph{In IIB supergravity ~\cite{Green82}   :} A is a four form 
that is self dual ($\hat{H} = \, ^{\ast } \hat{H} $) and there are
a  scalar $\chi $ and another two form $ B' _{\mu \nu } $ with strength 
$H' = \, dB'$. $^{\ast }$ denotes the Hodge duality operator. 

This mirrors the properties of the superstring theories that allow for
D-p-branes if $p$ is even/odd in IIA/B respectively. To see what fields
are allowed it is best to study the vertex operators ($S$ a Ramond
spinor) $ V = S^{\alpha}_{left} [CH_{\mu _{1} , \cdots , \mu _{p+2}}
\gamma ^{\mu _{1} , \cdots , \mu _{p+2}} ] _{\alpha \beta } S^{\beta }_{right} $ 
where $H$ again the field strength and $C \gamma ^{\mu} C^{-1} = -(\gamma ^{\mu} )^{T}
$. Together with the boundary conditions for the spinors ~\cite{POL96} 
\begin{eqnarray}
S_{R_{(z)}} = \pm  \Gamma ^{0} \cdots \Gamma ^{p} S_{L_{(\bar{z})}}  \label{SgammaS}
\end{eqnarray}
it follows that same/opposite $\gamma ^{\mu _{1} , \cdots , \mu _{p+2}} $ 
-parity implies odd/even rank of H and a very similar condition for the
supersymmetry parameters that generate the unbroken supersymmetries
follows from equation (\ref{SgammaS}) ~\cite{Malda96}:
\begin{eqnarray}
\epsilon _{R} = \Gamma ^{0} \cdots \Gamma ^{p} \epsilon _{L} \label{egammase}
\end{eqnarray}
which is only possible for even/odd $p$ in IIA/B. We discussed
the S-self-duality of IIB superstring theory and again we find that the
IIB supergravity still has this symmetry. Because of the S-duality being
the duality $\phi \leftrightarrow - \phi $ it is convenient not to use the
string but the S-duality invariant Einstein metric (equation (\ref{EM})). It follows: 
\begin{eqnarray}
S_{II} = ( \frac{1}{16 \pi G^{(10)}_{N}}) \int \, d^{10} x \, \sqrt{-g_{E}} \,
[ (R_{E} + 4(\bigtriangledown  \phi )^{2} - 
\frac{1}{3} e^{-8\phi /(D-2)}  H^2 ) + \ldots \, ] 
\end{eqnarray}
 
\subsection{Ramond-Ramond Solitons}
In the present work I first described M-theory and its D-branes and then
supergravity. Historically there was first the effective supergravity and then the
question: What are the solitons (localised field energy stabilized by
topological charges) ? Because we find p+1 forms in the RR sector and
tensor potentials couple to extended objects, they must be p-branes.
Form-theoretically one writes in (d-1)+1 dimensions of the spacetime for
the point charge (electric) and the dual point charge (magnetic) the
generic Gauss law:
\begin{eqnarray}
Q^{el} = \int _{R_{d-1}} \, ^{\ast} j =  \int _{R_{d-1}} \, d^{\ast} F =
\int _{S_{d-2}} \, ^{\ast} F\\ 
Q^{mag} = \int _{R_{d-1}}  j =  \int _{R_{d-1}} \, dF =
\int _{S_{d-2}} \, F 
\end{eqnarray}
in order to express that the charge is measured by the long range
fields. $^{\ast}$ is the Hodge duality operator, $S_{d-2} $
is at spatial infinity and j is a (d-1) form current. In quantum
electro dynamics j is a three form because the electron is a point particle.
If the p-brane is not compactified to a point but an infinite $q$-hyperplane than $(d-2)
\rightarrow (d-q-2)$ for the electric brane and so on.

$Q^{el}$ and $Q^{mag}$ should obey the Dirac quantisation condition
($Q^{el} Q^{mag} = \pi n \, ; \, n$ is an integer) in the quantised theory. The
coupling is
\begin{eqnarray}
\mu _{p} \int _{R_{p+1}} \, A_{p+1} = \mu _{p} \int A_{\mu _{1} , \cdots , \mu _{p+1}}
(\frac{\partial x^{\mu _{1}}  }{\partial \sigma ^{1  }} ) \cdots
(\frac{\partial x^{\mu _{p+1}}}{\partial \sigma ^{p+1}} ) \,    d^{p+1}
\sigma 
\end{eqnarray}
for the electric p-brane; for example 
\begin{eqnarray}
e\int A_{\mu} (\frac{\partial x^{\mu }}{\partial \tau} ) \, d\tau = j A 
\end{eqnarray}
in quantum electro dynamics. The magnetic fields are not necessary but
expected because one expects the generic electric charge to be
quantised. Such a (electric) charge quantisation is either understood to be
topological --- then one has most certainly as well topological arrangements that
are magnetic --- or it is only that one would like at least one magnetic
charge that leads then to quantisation of the electric charges via the
Dirac quantisation condition which is here the only quantisation due to
quantum mechanics. In 10 dimensions the magnetic dual charge is due to a
(6-p)-brane (D-4-p) that couples to a $A' _{7-p} $ form related through
equations like  $dA' _{7-p} = \, ^{\ast} dA_{p+1} $ such that for $F$
and $^{\ast}F$ holds $[(7-p)+1]+[(p+1)+1]=10$. In perturbative string
theory there are no fundamental objects (perturbative string states)
that have charges under the described fields.
Therefore, both, the electric and the dual magnetic charges must
be solitons and indeed the p-branes that are BPS turn out to be
D-p-branes. More precisely: the D-branes we encountered in the
non-perturbative  superstring theory (i.e.: via string dualities) and
that are described by conformal field theories turn out to be extremal
p-branes in the supergravities that we obtained from the superstring
theories. The D-p-brane tension goes as $\frac{1}{(\alpha ')^{(p+1)/2} g }$
(equation (\ref{tensions})), where the tension is the
 mass per spatial volume of the D-p-brane just
like for the D-string ($mass/length = T = 1/(2\pi g \alpha ')$)
(equation \ref{tensions} etc.). To summarize:
The S-duality (electric-magnetic duality) leads to the following pairing of
D-p- and D-(6-p)-branes: In IIA

\begin{itemize}
\item $ A_{\mu}				$ :	electric 0-brane and magnetic 6-brane
\item $ C_{\mu \nu \rho}		$ : 	electric 2-brane and magnetic 4-brane
\end{itemize}
and in IIB which has the zero form $\chi $ as well
\begin{itemize}
\item $ \chi				$ : 	electric (-1)-brane (the
				instanton in the Euclidean theory) and magnetic 7-brane
\item $ B'_{\mu \nu}			$ :	electric 1-brane and magnetic 5-brane
\item $ A_{\mu \nu \rho \sigma}$ : 	self dual 3-brane
\end{itemize}
 
Of course, the above is the RR sector. There are other quite similar
pairs like the important 1-brane (fundamental string) paired with the
solitonic 5-brane (that is 1 dual to 5 under $B$ instead of $B'$):
\begin{itemize}
\item 1-brane: worldsheet on $(x^{0},x^{9})$ 		which is $\perp
R_{8} \, $ with $\partial R_{8} = S_{7}$
\item 5-brane: worldvolume on $(x^{0}, \cdots ,x^{5})$ 	which is $\perp
R_{4} \, $ with $\partial R_{4} =S_{3}$
\end{itemize}
with $3+7=10$ , tension $ T \propto Q^{el} = \int _{R_{8}} \, ^{\ast}
j^{el} = \int _{S_{7}} e^{-\phi} \, ^{\ast} H$ , 
$Q^{mag} = \int _{R_{4}} j^{mag} = \int _{S_{3}} H$ and Dirac
quantisation  $(Q^{el} Q^{mag} = \pi n)$ --- just to stress that the
duality relations do not only apply to the RR-sector. This $(1,5)$-pair
is important because we can calculate the gravitational constant with
help of the Dirac quantisation for this pair. With equations
(\ref{Tfund}), (\ref{Ts5}), (\ref{Gd}) and the Dirac quantisation
condition we can write immediatelly
\begin{eqnarray}
G_{N}^{(d)}  M_{string} M_{5-brane} \propto \frac{G_{N}^{(10)}}{V_{T}}
\frac{R_{9}}{\alpha '} \frac{R_{9} \cdots R_{5}}{(\alpha ')^3 g^2} = n
\pi \\
\Rightarrow G_{N}^{(10)} \propto g^2 (\alpha ')^4  
\end{eqnarray}    
which is equation (\ref{Gg}). Considering the various factors of $2 \pi
$ and demanding that there is no singularity like the Dirac string that
appears when we introduce a magnetic monopole into the quantised
Maxwell theory results in 
\begin{eqnarray}
G_{N}^{(10)} = 8 \pi ^6 g^2 (\alpha ')^4
\end{eqnarray}

\subsection{Matching D-brane Configurations with the Relevant Supergravities}
\subsubsection{Introduction}
The aim is to obtain thermodynamics from a microstate description. The
strategy  will be simple: Compare supergravity solutions with the
 BPS string/brane configurations that have the same quantum numbers
(charges). Especially compare whether their degeneracy due to winding charges etc.
fits to the entropy of the area of the event horizon
that comes from the metric of the supergravity solution. 

We have to clearly distinguish between the two sides of it. One is the effective
supergravity side; i.e. the finding of BPS black holes with non-zero
area of the event horizon.  Then we calculate $S$, $T_{H}$ and so
on and this gives the thermodynamics of the black hole due to
superstring theory (namely its low and perturbative energy approximation) but not yet the
microstate description. The other side is the identification of D-brane
configurations and the calculation of the emerging thermodynamics from
the statistical mechanics of the microstates. If we find agreement with
the results from the supergravity side then this will boost our
confidence in the consistency of superstring theory (M-theory). Thus,
the relatively new thing is not that superstring theory describes black
holes (thats ``old'' apart from the fact that the superstring inspired
gravities have charges not carried by the fundamental string and need a
D-brane interpretation). New is that M-theory gives a microstate
description. In the light of the recent successes of the D-brane picture
I will present it first and after that the supergravity side because
this facilitates the unterstanding of the latter.

\subsubsection{The D-brane Side of the Matching}
There are at least two reasons why one needs more than just one black
brane in order
to model a black hole --- and we want a black hole because the
Bekenstein-Hawking entropy law is derived for black holes and not for
just any body and its would-be event horizon if it were compressed
enough. The supergravity solution of a D-brane has a Schwarzschild
radius of order $r^{7-p} \sim g \ll \sqrt{\alpha '}$. Thus, any strings
on it or interacting with it will not see the event horizon. This is
obviously not a good model for black hole.
Secondly, using only one type of brane leads
to scalar fields that diverge at the event horizon. Therefore, the event
horizon is not smooth and the black hole is not like the ones we have in
general relativity with non-zero entropy. The latter is what we would
like to model. Thus we need superpositions of many branes for a large
Schwarzschild radius and moreover different types of branes so that
scalars of different type may balance each other at the horizon. They
can do so because scalars (like the dilaton determining the size of
$R_{11}$ ) are pressures and tensions in the direction of compactified
dimensions. For example the dilaton field generated by $Q_{p}$
coincident p-branes in $D=10$ dimensions is
\begin{eqnarray}
e^{\pm 2 (\phi _{10} - \phi _{\infty })} = f_{p}^{\frac{p-3}{2}} \\
f_{p} = f_{p}(x,Q_{p})  
\end{eqnarray}  
with $(+)$ for the NS branes, $(-)$ for the D-branes and $f_{p}$ will be
discussed a bit closer later on. Thus, with different kinds of branes
and appropriate choices of the charges $Q_{p}$ one may balance scalars
like the dilaton for all $x$.  

Having different kinds of D-branes leads to more boundaries
and related constraints on the spinors due to equation (\ref{egammase})
being valid for different p simultaneously. Investigation of these   
constraints leads to information about whether and how many      
supersymmetries survive (recall that the $\epsilon $ in equation
(\ref{egammase}) are the parameters generating the unbroken
supersymmetries) and to the BPS-bound (mass formula) due to the
dimension p of the involved D-branes and due to their orientation,
i.e. whether they are parallel or  orthogonal or it might be a 
tilted intersection between them. For example $\Delta p $ (that is 
$(p-p')$ with $p'$ the dimension of another brane)
being 4 or 8 preserves one quarter of the supersymmetries in the
parallel case ~\cite{Douglas95}.

In order to have winding charges other than the winding around the
11th dimension we need to compactify dimensions (apart from the
incentive to be left with a $3+1$-spacetime we would like to play with 
charges and balance scalars). The simplest way for
spacetime and D-branes is a compactification on a torus $T^{p}$ that   
gives an identification on a lattice to the spacetime
\begin{eqnarray}
x_{\mu} \equiv x_{\mu} + n_{s} a_{\mu} ^{s}\\
\mu = (10-p), \cdots , 9 \\
n_{s} \in Z
\end{eqnarray}
where the $a_{\mu} ^{s} $ are the lattice basis vectors.
This compactification means a periodicity for the properties of the  
D-p-brane along the directions of the lattice basis vectors which is 
ensured for the D-p-brane via its translation/boost invariance
only for directions parallel to the brane like in the case where we
``wrap'' the p-brane around the compactification torus $T^{p}$.
Orthogonal to the branes one needs to
superimpose D-branes in order to create a periodic function along the  
compactified directions which means a lattice of branes on the lattice
of compactification.

\subsubsection{The Supergravity Side}
In order to study the appropriate
solution of the supergravity we need not write down the whole and
complex supersymmetric action and we do not need to solve the second
order Euler Lagrange equations. It is sufficient to work with an ansatz
for the metric that is quite general but has the
symmetry one expects the black hole to have; $SO(1,p) \times  SO(9-p)$
for example in case of a single D-p-brane in the 10 dimensional string
frame gives the following ansatz for the line element:
\begin{eqnarray}
(d\tau )_{string}^2 = (\frac{1}{\sqrt{f_{p}}}) [(dx^{0})^2 -
 d\vec{x} _{\parallel} \, d\vec{x} _{\parallel} ] -
 (\sqrt{f_{p}} ) \, [ d\vec{x} _{\perp} \,d\vec{x} _{\perp} ]
\label{lineelementoneDp} \\
\vec{x} _{\parallel} = (x_{10-1}, \cdots , x_{10-p} ) \\ 
\vec{x} _{\perp} = (x_{1}, \cdots , x_{9-p} ) 
\end{eqnarray}
Then one has to consider only the first order supersymmetry equations that set
the dilatino and gravitino variations to zero
\begin{eqnarray}
\delta \Psi \, = \, \delta \lambda \, = 0
\end{eqnarray}
because we work with BPS states and there are preserved supersymmetries.
Allowing for the fields we want to keep (like the dilaton and the RR
field B) and setting the other fields to zero one obtains conditions
relating metric and fields. Equipped with these conditions on the ansatz
on one hand and the dilatino and gravitino variations being zero on the
other one obtains constraints like equation (\ref{egammase}) again. 

Note that parallel to the relations mass $\propto 1, 1/g$ and $1/g^2 $
for the fundamental string, the D-branes and the NS solitons (solitonic
5-brane) respectively one obtains:  
\begin{eqnarray}
(d\tau _{D-p-brane}  )^2 = \sqrt{f_{p}} (d\tau _{fundamental} )^2 \\ 
(d\tau _{NS-p-brane} )^2 = \sqrt{f_{p}} (d\tau _{D-p-brane}   )^2
\label{tauf}  
\end{eqnarray}
For a concrete example consider the long fundamental string 
(not D-string). The $SO(1,1) \times SO(8)$ ansatz is 
\begin{eqnarray} 
(d\tau )^2 \propto  (f_{p}^{-1}) [(dx^{0})^2 - (dx^{9})^2 ] - 
[ d\vec{x} _{\perp} \,d\vec{x} _{\perp} ] \\
\vec{x} _{\perp} = (x_{1}, \cdots , x_{8} ) 
\end{eqnarray}
and one allows the dilaton and $B_{09}$ to be excited. In type IIA we
require ~\cite{Malda96} 
\begin{eqnarray} 
\delta \lambda 		\, = [\partial _{\mu} \phi \gamma ^{\mu} \Gamma _{11}
+ (\frac{1}{6} ) (H_{\mu \nu \rho} \gamma ^{\mu \nu \rho}        ) 		] \eta = 0 \\
\delta \Psi _{\mu }	\, = [\partial _{\mu} 	
+ (\frac{1}{4} ) (\omega _{\mu}^{ab} + H_{\mu}^{ab} \Gamma _{11} )  \Gamma _{ab}] \eta = 0
\end{eqnarray}
where $\eta = \epsilon _{R} + \epsilon _{L} $ is the sum of one spinor
with positive and one with negative chirality. $\omega _{\mu}^{ab}$ is
the spin connection corresponding to the zehnbeins $e_{\mu}^{a}$ and the $\Gamma$ 
satisfy the Clifford algebra $\{ \Gamma _{a} , \Gamma _{b} \} = 2 \eta
_{ab} $ with $\gamma ^{\mu} = e_{a}^{\mu} \Gamma ^{a} $.

These relations lead to the following modified ansatz ~\cite{Dab8990}
that keeps half of the supersymmetries:
\begin{eqnarray}
(d\tau )^2 = (f^{-1}) [(dx^{0})^2 - (dx^{9})^2 ] -
[ d\vec{x} _{\perp} \,d\vec{x} _{\perp} ] \label{lineelementfundamental} \\
B_{09} = 1/2 \, (f^{-1} - 1) \label{B09} \\
f = e^{-2(\phi - \phi _{\infty })} \, \, ; \, \phi = \phi _{(\vec{x_{\perp}})} 
\end{eqnarray} 
which gives with $H = \, dB$: 
\begin{eqnarray}
H_{09a} = \partial _{a} \phi e^{2(\phi - \phi _{\infty})}
\end{eqnarray} 
and $\delta \Psi _{\mu} \, = \, \delta \lambda \, = 0 $ will be satisfied
when (~\cite{Malda96} equation (2.6)):
\begin{eqnarray}
\epsilon _{R/L} = (f^{-1/4}) \epsilon _{R/L}\\
\Gamma ^{0} \Gamma ^{9}  \epsilon _{R/L} ^{0}= \pm \epsilon _{R/L} ^{0} 
\end{eqnarray}
where $ \epsilon _{R/L} ^{0} $ are asymptotic values of the Killing
spinors which generate the infinitesimal local supersymmetry
transformations. The dilaton's
$ \vec{x_{\perp }} $-dependence is still arbitrary and the
equation of motions only require $f$ to be a harmonic function of
$\vec{x_{\perp }} $
\begin{eqnarray}
\vec{\partial _{\perp }} \vec{\partial
_{\perp }} f = 0
\end{eqnarray}
In order to model a black hole we might like to put 
\begin{eqnarray}
f = (1+ \frac{Q}{r^{6}})
\end{eqnarray}
(after compactification of one dimension $r^{d-3} = r^{9-3} = r^{6}$) which
gives a curvature singularity and the event horizon at $r=0$.

Note that in general the whole apparatus of general relativity as
discussed  can now be applied.  That is, for a
solution that is Minkowski flat at spatial infinity we may expand the
Einstein metric ($g_{E_{\mu \nu}}=\eta _{\mu \nu } + h_{\mu \nu }$ where
 $ h_{\mu \nu }$ is small ) in order to calculate the ADM mass with
surface integrals at infinity. We may read off charges, angular
momentum, Bekenstein-Hawking entropy and Hawking temperature
straightforwardly from the metric.
 
For simplicity the examples above are in the D=10 spacetime but the
macroscopic observer ($d < D$) who does not see the compactified dimensions will
request a purely d-dimensional supergravity theory. Therefore we have to do a
dimensional reduction: \emph{reduction} on the supergravity side and
\emph{compactification} on the superstring theory side. The ansatz is
the same with $d$ substituting $D$ but after this Kaluza Klein
dimensional reduction we have to consider new fields when determining
the solution. These fields are Kaluza Klein gauge fields and Kaluza
Klein scalars.  Considering all scalars the result is that if only a
single type of RR charge is used (that corresponds to a single 
type of D-brane), the event horizon will have zero area and the solution
is singular at the event horizon because of unbalanced and diverging scalars.

\section{The Black Hole Microstate Descriptions}
\subsection{Introduction}
There have been numerous early efforts to describe black hole entropy
via microstates; as early as 1975 ~\cite{Bekenstein75}. That only
superstring theory as the only consistent quantum theory of gravity is
the right place to look for such microstate descriptions has been
stressed by t`Hooft (for example in ~\cite{Hooft9091}) and later very much
by Susskind who suggested that very massive string states should be
black ~\cite{SUSK9394} . During that time one still thought that black
holes must be fundamental string states but this clashes  with the
fact that the logarithm of the degeneracy of elementary string states is
proportional to the mass $M$ but the Bekenstein-Hawking entropy is
proportional to $M^2$ (see equation (\ref{M})). ~\cite{SUSK9394} and 
~\cite{Russo94} suggested that $M^2 _{black \, hole} = M_{string}$ 
because of large mass renormalisations. This is one more reason to
study BPS states because they do not receive renormalisation corrections.

More recent approaches to black hole microstates have been attempted
notably by ~\cite{Sen95} ~\cite{Carlip95} ~\cite{Larsen95}
~\cite{Cvetic95} ~\cite{Tseylin96}. Although they proposed a
quite different string theoretical approach (different from the now so
widely known approach starting with ~\cite{S+V96}) it is remarkable that
~\cite{Larsen95} anticipated the results of Strominger and Vafa rather
closely and on general considerations in the framework of superstring
theory. The Strominger-Vafa computation ~\cite{S+V96} was the first
success in accounting for the Bekenstein-Hawking entropy via
microstates, but I will not start with their work because there are now
equivalent but easier ways. What will be historically correct though is to
start with IIB theory compactified on $T^{5}$ as ~\cite{S+V96}
have done because here the analysis is still the simplest of the ones
found yet. 

In the next sections we will see that the string theoretical
microstate-descriptions all suggest that the evolution of a black
hole is unitary. We might not trust the results in the strong coupling
regime because the D-brane description is not valid there and we have to
rely on the results being protected by supersymmetry due to the BPS
nature of the configurations. But on the D-brane side of the comparison
between D-brane configuration and supergravity effective theory one has
a very clear \emph{topological} picture with the entropy coming from the
degeneracy of possible numbers of branes and possible windings ---
especially for BPS states that have $nm=0$ where n is the momentum
quantum number and m counts the winding because T-duality is
$(n \leftrightarrow m)$-duality. Cranking up the coupling there is no
reason to assume this topological picture will change as long as one
does not encounter phase transitions.

\subsection{Extremal and Near-Extremal Black Holes in 4 and 5
Dimensions}
\subsubsection{Extremal d=4+1 and d=3+1 Black Holes}
For the following I go more or less closely along with the
argumentations of \\ ~\cite{Callan96} for $d=4+1$, ~\cite{MaldaStrom96}
and ~\cite{Johnson96} for $d=3+1$  and ~\cite{Malda96} for both. I tried
to add with notations and explanations that I as a beginner would have
liked to find.
The literature reflects the anxiety that accompanies the pressure to
publish and the swift pace with which the subject is moving ahead. I
changed almost everything in order to be left with a convenient and
consistent notation rather than with a historically correct account.
The result is that the arguments look very straightforward to a degree
that one might wonder why the early work did not derive it that way.\\

The D-1-brane is the simplest D-brane we can wrap in different ways in
order to have degeneracy and therefore entropy due to windings for
example. In $D=10$ a D-p-brane produces the dilaton field
\begin{eqnarray}
e^{-2\phi_{(10)} } = f_{p}^{\frac{p-3}{2} } \label{dilatonofpbrane} 
\end{eqnarray} 
where we assumed $\phi _{\infty } = 0$ from now onwards since that is
how we expect fields to behave when they are far from the black hole. 
In order to have solutions that do not blow up at the event horizon one
has to balance the dilaton. A D-5-brane would be very convenient because
the matching powers of the $f_{(x)}$-functions make them balanced for all
$x$ if they are for any one $x$: 
\begin{eqnarray}
e^{-2\phi_{(10)} } = f_{1}^{\frac{1-3}{2} } \times  f_{5}^{\frac{5-3}{2} } 
= f_{1}^{-1} \times f_{5}^{+1} 
\end{eqnarray}
In order to wrap away a 5-brane (we want to model a spherically symmetric
black hole) we need a 5-dimensional torus of compactification.
D-1 and D-5 branes are S-dual in the IIB theory. Thus if we take the IIB
theory then we need only consider one field, namely the RR 2-form $B'$.
Therefore consider IIB superstring theory on $T^{5}$. The torus is
\begin{eqnarray}
T^{(D-d)} = S^{1}_{(10-1)} \times S^{1}_{(10-2)} \times \cdots \times 
 S^{1}_{(d)}  
\end{eqnarray}
with ``volume'' (volume$/(2 \pi )^{(D-d)}$) 
\begin{eqnarray}
V=\prod_{\mu = d}^{9} R_{\mu } 
\end{eqnarray}
so that a $d=(d-1)+1$ uncompactified spacetime is left and has the
familiar indices $\mu = 0,1,\ldots ,(d-1)$.
A $T^{(D-d)}$-compactification is the simplest one but does not yield
the physics of our world. We would like to have $d=3+1$ though to be a
bit closer to our world and therefore we consider a theory on $T^{6}$.
Because IIB on $T^{5}$ turned out successful, the recipe is: Take the
$d=4+1$ configurations and compactify one more direction by doing a
T-duality transformation along $x_{\mu = 3+1}$ and having a lattice of
the $d=4+1$-configurations on the lattice of compactification. 
This is exactly the way in that we compactified D-p-branes earlier on. 
The T-duality transformation has two effects: The D-1- and D-5-branes
become D-2- and D-6-branes respectively ($\Rightarrow$ the dilaton will
not be balanced anymore) and the IIB theory on $T^{5}$ becomes a IIA
theory on $T^{6}$ because IIB and IIA are related by T-duality as we
have discussed. Compactifying like this leaves a relatively simple model
in that the branes of highest dimensionality have only one way to be
wrapped away (D-6-brane on $T^{6}$). In order to balance the scalar (equation  
(\ref{dilatonofpbrane})) we take (NS) solitonic 5-branes ($p=``s5$'')
because IIA has not got suitable RR-5-branes (D-5-branes):
\begin{eqnarray}
e^{-2\phi_{(10)} } = f_{2}^{\frac{-1}{2} } f_{6}^{\frac{3}{2} } 
f_{s5}^{\frac{-2}{2} } \label{balanced}
\end{eqnarray}
where the dilaton field of a NS p-brane is 
\begin{eqnarray}
e^{+2\phi_{(10)} } = f_{p}^{\frac{p-3}{2} } 
\end{eqnarray}

There are now 27 charges due to the gauge fields (i.e. D-branes
etc.) possible on the torus $T^{5}$ for example: 5 momentums, 5 D-string windings, 5
string windings, $1+2+3+4=10$ D-3-brane wrappings, 1 D-5-brane and  1 solitonic
5-brane wrapping. By ``$N$ wrappings'' is meant that there are $N$
choices of the set of internal directions along which the entire brane
is wrapped away. Thus there are $2 \times 27 = 54 $ charges together with
the anti charges. An anti brane is a brane with opposite orientation
and anti momentum is right instead of left moving. The variables $Q_{p}$
are integers and count the number of units $p$-charge. We will only be
concerned with $p=1,2,5,6$ for D-p-branes, $p=``N$'' (i.e. $Q_{N}$) for
units of momentum and $p=``s5$'' for solitonic 5-branes.\\

\emph{IIB on $T^{5}$ for $d=4+1$:}
\begin{itemize}
	\item $Q_{1}$ D-strings  wrapped along $S^{1}_{(10-1)}$
	\item $Q_{5}$ D-5-branes wrapped on $T^{5}$
 	\item $Q_{N}$ units of momentum $P_{(10-1)} =
		\frac{1}{R_{(10-1)}} $
		along  $S^{1}_{(10-1)}$ in one direction only (say left).
\end{itemize}
This is the S-dual to the solution of ~\cite{Cvetic95} ~\cite{Tseylin96}
because fundamental strings and solitonic 5-branes are S-duals to D-1-
and D-5-branes.\\

\emph{IIA on $T^{6}$ for $d=3+1$:}
\begin{itemize}
        \item $Q_{2}$ D-2-branes  wrapped on $S^{1}_{(10-1)} \times S^{1}_{(10-2)}$
        \item $Q_{6}$ D-6-branes  wrapped on $T^{6}$
        \item $Q_{N}$ units of momentum $P_{(10-1)} =
		\frac{1}{R_{(10-1)}} $ 
		along  $S^{1}_{(10-1)}$ in one direction only (say left).
	\item $Q_{s5}$ 5-branes wrapped on $S^{1}_{(10-1)} \times
		S^{1}_{(10-3)} \times \cdots \times 
		S^{1}_{(10-6)}$ (not on $ S^{1}_{(10-2)}$)
\end{itemize}
The equations given by formula (\ref{egammase}) applied to all the
D-branes and momentums show that the solitonic 5-brane does not break
additional supersymmetries. If $Q_{s5}=0$ this arrangement will be
T-dual to the one for IIB on $T^{5}$.

In the case that there are only branes ($Q_{N}=0$), the ansatz on the
supergravity side is straight forward because the metric is a ``local
thing'' and the effects on it simply add up linearly. Thus with
equations (\ref{lineelementoneDp}) and (\ref{tauf}) we obtain in the
$D=10$ dimensional string frame:
\begin{eqnarray}
(d\tau )_{string}^2 = \sum_{p-branes} (d\tau _{p})_{string}^{2} \\ 
=  \sum_{D-p-} (\frac{1}{\sqrt{f_{p}} })
 	[(dx_{0})^2 - (d\vec{x} _{\parallel p} )^2 ] -
 (\sqrt{f_{p}} ) \, [ (d\vec{x} _{\perp p} )^2 ] \nonumber \\
+  \sum_{sp-} [(dx_{0})^2 - (d\vec{x} _{\parallel p} )^2 ] -
  	f_{p} \, [ (d\vec{x} _{\perp p} )^2 ]
\label{tausumtaus} 
\end{eqnarray}
This becomes especially simple because the intersections of branes are
all taken as orthogonal or parallel and we consider only $ (p,p') =
(1,5) $
or $ (2,6) $ with the $ (p < p') $-branes lying inside the $p' $-branes. With
$ Q_{s5} = 0 $ for example:

\begin{eqnarray}
(d\tau )_{string}^2 = (\frac{1}{\sqrt{f_{p}} }) (\frac{1}{\sqrt{f_{p+4}} }) 
        [(dx_{0})^2 - (d\vec{x} _{\parallel } )^2 ] \nonumber \\
	- (\sqrt{f_{p}} )  (\frac{1}{\sqrt{f_{p+4}} })
	[ (d\vec{x} _{\perp \parallel } )^2 ] \nonumber \\
	- (\sqrt{f_{p}} )  (\sqrt{f_{p+4}} )  
	[ (d\vec{x} _{\perp } )^2 ]
\label{tausumperpar}
\end{eqnarray}
where those directions which are parrallel/orthogonal to all branes are
denoted by
\begin{eqnarray}
 \vec{x} _{\parallel } = (0, \ldots , 0, x_{10-p}, \ldots , x_{9})  \\
 \vec{x} _{\perp } = (x_{1}, x_{2}, \ldots , x_{d-1}, 0, \ldots ,0 )
\label{shitvectors}
\end{eqnarray}
and orthogonal to the D-p-brane but parallel to the D-(p+4)-brane are
\begin{eqnarray}
 \vec{x} _{\perp \parallel } = (0, \ldots , 0, x_{d}, \ldots , x_{9-p}, 0,
\ldots , 0)  \label{vector} 
\end{eqnarray}
The dilaton field of this ansatz is (equation (\ref{dilatonofpbrane}))
\begin{eqnarray}
e^{-2\phi_{(10)} } = f_{p}^{\frac{p-3}{2} } f_{p+4}^{\frac{p+1}{2} }
= \frac{f_{5}}{f_{1}} \, \, \, or \, \, \sqrt{\frac{f_{6}^{3}}{f_{2}} }
\label{unbalanced}
\end{eqnarray}
and the harmonic functions $f$ turn out to be dependent on the d-spacetime
radius $r$:
\begin{eqnarray}
r^{2} = (\vec{x} _{\perp } )^{2} \\
 f_{p} = 1 + \frac{c_{p}^{(d)} Q_{p}}{r^{(d-3)}} 
\label{dependence}
\end{eqnarray}  
where the power of $r$ is dependent on $d$ because any further
compactification needs a lattice structure of the configuration we had
before the compactification. Therefore we obtain other harmonic
functions $f$ whose dependence on $r^{(d-3)}$ reflects that the large
scale observer does not observe any internal dimensions.
The $c$-proportionality factors are due to rather tedious calculations
involving the mass of the D-branes and the gravitational constant in
$D=10$ dimensions ~\cite{Malda96}, for example:   

\begin{eqnarray}
c^{(5)}_{1} =  \frac{4 G_{N}^{(5)} R_{9}}{\pi \alpha ' g} \label{c} \\
c^{(5)}_{5} =  \alpha ' g   \label{c5}  
\end{eqnarray}        

The supersymmetry breaking leads to (use equation (\ref{egammase})):
\begin{eqnarray}
 \epsilon _{R} = \Gamma ^{0} \Gamma ^{9} \epsilon _{L} \\
 \epsilon _{R} = \Gamma ^{0} \Gamma ^{5} \ldots \Gamma^{9} \epsilon _{L} 
\label{epsilonsagain}
\end{eqnarray}
for $(p,p') = (1,5)$ for example. In order to balance the dilaton in the
IIA on $T^{6}$ case we need $Q_{s5} \neq 0$ which gives with  equations
(\ref{tausumtaus}) and (\ref{tausumperpar}) 
\begin{eqnarray}
(d\tau )_{string}^2 = (\frac{1}{\sqrt{f_{2}} }) (\frac{1}{\sqrt{f_{6}} }) (1)
        [(dx_{0})^2 - (dx_{9} )^2 ] \nonumber \\
	-(\frac{1}{\sqrt{f_{2}} }) (\frac{1}{\sqrt{f_{6}} }) (f_{s5})
	[(dx_{8})^2 ] \nonumber \\
        - (\sqrt{f_{2}} )  (\frac{1}{\sqrt{f_{6}} }) (1)
        [ (dx_{4})^2 + \cdots + (dx_{7})^2] \nonumber \\
        - (\sqrt{f_{2}} )  (\sqrt{f_{6}} ) (\sqrt{f_{s5}} )
        [ (d\vec{x} _{\perp } )^2 ]
\label{4terms}
\end{eqnarray}
and with equation (\ref{balanced})
\begin{eqnarray}
e^{-2\phi_{(10)} }= \sqrt{\frac{f_{6}^{3}}{f_{2}^{1} f_{s5}^{2}} }
\end{eqnarray}
, thus the dilaton is balanced if we chose the charges appropriately. 
For the fields one obtains: \\
\emph{IIB on $T^{5}$}:
\begin{eqnarray}
B_{09} = 1/2 \, (f_{1}^{-1} - 1) \\
H'_{ijk} = 1/2 \, \epsilon _{ijkl} \partial _{l} f_{5}
\end{eqnarray}
~\cite{Malda96}, compare with equations (\ref{B09}) etc. \\

\emph{IIA on $T^{6}$}:
\begin{eqnarray}
C_{049} = 1/2 \, (f_{2}^{-1} - 1) \\        
H_{ij4} = 1/2 \, \epsilon _{ijk} \partial _{k} f_{s5} \\
(dA)_{ij} = 1/2 \, \epsilon _{ijk} \partial _{k} f_{6}
\end{eqnarray}
~\cite{Cvetic950} ~\cite{Tseylin96}, where the $\epsilon $-tensor
density is the $\epsilon $-tensor for the
flat $\vec{x_{\perp }} $-space (the flat space volume form) and
$i,j,k,l \in \{1, 2, \ldots , (d-1) \} $. 

The brane configurations as
they are up to now ($Q_{N} = 0$) are put together 
in a way that balances the dilaton fields of the branes. But gravity
curves spacetime and that applies as well to the internal compactified space. 
Thus, actually the balancing is as well a compensation of brane tension
parallel to the branes with help of the pressure due to the electric fields
perpendicular to other branes. These forces determine the internal metric
and the compactification radii $R_{\mu }$. Thus, the dilaton is just one
of the scalars we would like to balance in order to model the black
holes that we are used to in the context of general relativity. 
All branes are parallel to $S^{1}_{9}$ and we need to put momentum along
this circle in order to have a stable $R_{9}$ as the compromise between the
tension parallel to the branes and the pressure of the momentum. Therefore we need
to put $Q_{N} \neq 0$ but these units of momentum do only go along
$S^{1}_{9}$ and into one direction only in order to have a solution that
preserves some supersymmetry and is still BPS. The complete  $Q_{N} \neq
0$-configuration owes its degeneracy to the different possible windings
of branes and to the momentum. The momentum we can interpret as
oscillations of the D-strings inside the D-5-branes  
(or D-2-branes and solitonic 5-branes inside the  D-6-branes) and as
moving open strings stretched between different branes. The oscillations are
parallel to the D-5(/6)-branes because the configuration is a bound
system. Thus the D-5(/6)-branes do not oscillate. The degeneracy is due
to all the ways that the $Q_{N}$ units of momentum can be assigned to
certain branes. The ansatz for the line element changes due to these
oscillations ~\cite{Callan91} ~\cite{Garf92}. This becomes a simple 
substitution in case we have only oscillations in internal dimensions
and no angular momentum ~\cite{Callan95} ~\cite{Malda96} : 
\begin{eqnarray}
[(dx_{0})^2 - (dx_{9})^2 ] \leftrightarrow [(dx_{0})^2 - (dx_{9})^2
+(1-f_{N}) (dx_{0} \, - \, dx_{9})^2 ] 
\label{simplesubstitution}
\end{eqnarray}
with the $c_{N}^{(d)}$ :
\begin{eqnarray}
   c_{N}^{(5)} =  \frac{4 G_{N}^{(5)} }{\pi R_{9}} \\ 
   c_{N}^{(4)} =  \frac{4 G_{N}^{(4)} }{R_{9}}  
\label{cN}
\end{eqnarray}
The supersymmetries that are broken by the addition of the momentum
states lead to formulas similar to (\ref{egammase}), here:
\begin{eqnarray}
 \epsilon _{R} = \Gamma ^{0} \Gamma ^{9} \epsilon _{R} \\
 \epsilon _{L} = \Gamma ^{0} \Gamma ^{9} \epsilon _{L}
\label{momentumepsilons}
\end{eqnarray}
An oscillating string preserves $1/4$ supersymmetries and our whole
configuration of two types of D-branes and momentum excitations
preserves (see equations (\ref{epsilonsagain}) and
(\ref{momentumepsilons})) only $1/2 \times  1/2 \times  1/2 = 1/8$, that
is 4 left from initially $2 \times (8_{R} + 8_{L}) = 32$ due to 8
transverse dimensions for every kind of right and left moving spinors.
The ansatz is still not entirely independent of the internal dimensions.
What we want is a $d$-dimensional supergravity with a metric that is only
dependent on $(dx_{0})$ and $(d\vec{x} _{\perp })$. Thus one needs to
perform the dimensional reduction. How one does this in detail one can
find in ~\cite{Maharana93} and  ~\cite{Sen93}. One finds in the
Einstein frame (using the Einstein metric):   
\begin{eqnarray}
(d\tau )_{E}^2 = (F^{-1})^{a_{(d)}} [(dx_{0})^2 ] - F [ (d\vec{x} _{\perp } )^2 ]
\label{tauF}  
\end{eqnarray}
(where $a_{(5)} = 2$ and $a_{(4)} = 1$) which is a d-dimensional black hole
because 
\begin{eqnarray}
F = [\prod_{p}f_{p} ]^{1/(d-2)} \label{F}  \\
\lim_{r \rightarrow 0}g_{\perp \perp } = \lim_{r \rightarrow 0}F = \frac{1}{r^{d-3}} 
[\prod_{p}c_{p} ]^{1/(d-2)}  [\prod_{p}Q_{p} ]^{1/(d-2)}  
\end{eqnarray}
thus the event horizon is at $r=0$ (this coordinate system does not
cover the inside of the black hole). In order to obtain a general Reissner-Nordstr\"{o}m
solution it is required that we do perform a certain balancing of the
dilaton. We need to set the charges into relation to each other. With
equation (\ref{lineelementd}) for the extremal black hole ($r_{+} =
r_{-}$) and equations (\ref{tauF}) and (\ref{F}) one finds that the
charges have to obey 
\begin{eqnarray}
r_{+}^{2} = r_{-}^{2} = c_{p} Q_{p} \label{cpQp}
\end{eqnarray}
for all $p$ in order to make the metric an extremal Reissner-Nordstr\"{o}m one. 
Note that the $r$ of equation (\ref{lineelementd}) is not the $r$ of the
equations (\ref{tauF}) and (\ref{F}) but a simple transformation will
show the equivalence. If (\ref{cpQp}) is satisfied the geometry of the
internal space will be independent of the uncompactified space
coordinates. This statement includes the dilaton since it is due to the
compactification of the 11th dimension. 

Now we may apply the formulas of general relativity. $T_{H}$ is zero
because the black hole is an extremal one. The entropy due to the area
of the event horizon $S = A^{(d)}_{+}/4G^{(d)}_{N}$ with $A_{+}$ being
a 3-dimensional ``area'' in the case of IIB on $T^{5}$, turns out to be 
\begin{eqnarray}
S = 2 \pi \prod_{p}\sqrt{Q_{p}} \label{SQ} 
\end{eqnarray}
where $p=1,5,N$ or $p=2,6,s5,N$. This equation reflects the topological
origin of the entropy of the BPS states because none of the continuous
parameters like $g,R$ or $V$ are in the equation. 

\subsubsection{Counting of States for the Extremal Black Holes}
Does equation (\ref{SQ}) fully agree with the D-brane picture, i.e. 
with the denumeration of
possible degenerate states ? ~\cite{S+V96} uses a very involved counting
based on Vafa's (~\cite{Vafa9511} and  ~\cite{Vafa9512} ) analysis of
the cohomology of moduli spaces for instantons. ~\cite{Callan96} have
given a slightly easier method of the counting but it is actually not
the ``simple'' denumeration and it will not possibly be if we want to
reproduce equation (\ref{SQ}) because it contains $\prod_{p}Q_{p} $ and
makes therefore no difference between many D-p-branes or a single
D-string wrapped $(Q_{1}Q_{5})$ times 
along $S^{1}_{9}$ --- as if D-1-branes and D-5-branes [or D-2-branes and
D-6-branes] are all mutually indistinguishable. 

All we know so far about the momentum is that there are $Q_{N}$ left
moving units of it on $S_{9}^{1}$. These units are carried by open
strings that are attached to the branes. This is the only way in that
one can describe D-brane momentum that goes parallel to the brane
because the D-brane is translation and boost invariant parallel to
itself. The oriented strings have start and end points that are attached
to the branes. Therefore, these open srings have Chan-Paton factors
$(i,f)$, for example $(5_{x},1_{y})$ for a string that starts on the
D-5-brane number $x$ and ends on the D-string number $y$ ($0<x<Q_{5}$
and $0<y<Q_{1}$). One possible way of the counting of states uses the
following simplifications:
\begin{itemize}
 
\item Although $f_{N}$ is evaluated by considering oscillations of
branes, i.e. moving massless and open strings attached to a brane, the
largest contributions to the entropy is due to open strings stretched
between different branes and even different kinds of branes like $(1,5)$
and $(5,1)$. The entropy due to $(1,1)$ and $(5,5)$ strings can be
neglected because although these open strings are massless, all open
strings considered together lead to interactions that are effectively
mass terms. The detailed analysis points to the $(i,f)$ strings with $i
\neq f$ as giving the configurations where the most constituents remain
massless. These will give the leading contributions to the entropy.
The latter is just the well known result about the populations of 
states of different energy in statistical mechanics.

\item Concentrating on $(i,f)$ strings with $i \neq f$, one finds that
for every possible numbers of
charges $Q_{p}$ the largest contributions to the entropy are due to the
windings that have the effective length $R_{eff} = (Q_{1} Q_{5} R_{9} )$
[ or $(Q_{2} Q_{6} R_{9} )$ ]. For example: If there is only one
brane of every type $p$ but the D-1-brane is wrapped
$(Q_{1} \times Q_{5})$
times around $S_{9}^{1}$. Or for another example:
There is only one D-string wrapped $Q_{1}$ times around $S_{9}^{1}$ and 
only one D-5-brane wrapped $Q_{5}$ times around $S_{9}^{1}$. The
Chan-Paton factors count the turns along $S_{9}^{1}$. If $Q_{1}$ and 
$Q_{5}$ have no common devisors a $(5_{x},1_{y})$ string will have to be
carried around $S_{9}^{1}$  for $(Q_{1} \times Q_{5})$ times until the
Chan-Paton factor is the same again [$(5_{x},1_{y}) 
\equiv (5_{x+Q_{5}},1_{y+Q_{1}})$].
The effective length is again $(Q_{1} Q_{5} R_{9} )$. Other examples are
more difficult to argue.\\

In general: The open strings are \emph{on the branes} and therefore
there are now $( Q_{p} Q_{p+4} Q_{N} )$ units of momentum
\begin{eqnarray}
\frac{1}{R_{eff}} = \frac{1}{R_{9} Q_{p} Q_{p+4}} \label{Reff}
\end{eqnarray}
where $p=1$ or $p=2$.
\end{itemize}

Both simplifications are only possible if the charges $Q_{p}$ are large
but eventually we want to have large charges anyway in order to have a
Schwarzschild radius that is bigger than the string length.  What now
remains is to count the possible states for the $(1,5)$ type string for example.
It has two orientations [$(1,5)$ and $(5,1)$] and can bind any of the
$Q_{1}$ D-strings to any of the $Q_{5}$ D-5-branes. In the case of
multiple winding the different turns of a brane contribute to the
counting. Thus we have ($2 Q_{1} Q_{5}$) possibilities. An analysis of
the Dirichlet and Neumann boundary conditions and the GSO (Gliozzi
Scherk Olive) chirality conditions\footnote{The GSO projection is
needed for the superstring in order to match the number of the bosonic
degrees of freedom to the number of fermionic ones. This truncation of
the spectrum removes the 
tachyonic ground state.} \cite{Callan96} for the fermions after that
leads to the result that the strings have only few more degrees of
freedom. In fact only two because they effectively live in  two
dimensions.  This gives ($4 Q_{1} Q_{5}$) possibilities for the
fermions. An analysis of the bosonic sector gives the same number.
The factor ($4 Q_{p} Q_{p+4}$) in the case of the IIA theory on $T^{6}$
may be derived in a similar way. The degrees of freedom for the
fermions are not decreased by the solitonic 5-branes because they do not
break additional supersymmetries. The entropy has a topological 
origin and  if $Q_{s5}$ is still zero IIA on $T^{6}$ will be the
T-dual to IIB on $T^{5}$. Thus we take the factor ($4 Q_{p} Q_{p+4}$) as
a general result.

The ``simple'' denumeration turns out rather complicated if we want to
follow it in detail and give a rigorous derivation. Nevertheless, there
are different interpretations of the configurations of charges possible
and therefore different countings. ~\cite{S+V96} started with $Q_{5}$
D-5-branes wrapped on $T^{5}$ and put $Q_{1}$ instanton solutions of 
the corresponding $U(Q_{5})$ gauge theory on $T^{4}$ of $T^{5} = T^{4}
\times S^{1}$. The instantons carry D-1-brane charge (one RR charge) but
are only D-1-branes after shrinking to zero thickness. As instantons
they can have different orientations and the factor ($4 Q_{1} Q_{5}$) 
is derived from the analysis of the moduli space for these instantons
~\cite{Vafa9511}~\cite{Vafa9512}.
One can equally well put $Q_{2}$ instanton solutions on the D-6-branes
and derive the factor ($4 Q_{2} Q_{6}$). In general one has to
investigate the moduli space 
\begin{eqnarray}
\frac{(T^{4})^{Q_{p} Q_{p+4}}}{\Sigma _{(Q_{p} Q_{p+4})}}
\end{eqnarray}  
where$\Sigma$ of $n$ is the permutation group of $n$ elements.
~\cite{HoroStrom96} simply argue that the $Q_{1}$ D-strings may wander
around inside the 4 transverse directions left for them inside the
D-5-brane. Therefore the number ($4 Q_{1} Q_{5}$) as the number of
massless bosons and the same number of fermionic supersymmetry partners. The
same goes for D-2-branes inside D-6-branes. In general, the D-p-branes
are \emph{bound} to the D-(p+4)-branes so that the oscillations 
can only be inside the 4 transverse directions inside the larger brane. 
~\cite{Johnson96} avoid a singular horizon and ensure a finite horizon
 area by having the D-brane configurations in the background of a
Kaluza Klein monople. Their $d=3+1$~(!) configuration has only the
three charges $(Q_{1}, Q_{5}, Q_{N})$ or the charges $(Q_{0}, Q_{4}, Q_{W})$
for the T-dual configuration where $W$ stands for the possible winding
of fundamental strings. Again the factor $(4 Q_{p} Q_{p+4})$ is found. 

For IIA on $T^{6}$ we add solitonic 5-branes and the factor 
($4 Q_{p} Q_{p+4}$) gets multiplicated by $Q_{s5}$ because the
solitonic 5-branes are distributed along $S_{10-2}^{1}$ which they
intersect. Therefore the open strings are further distinguished by the
pair of solitonic 5-branes they lie in between. Actually the intersected
D-2-branes break and end on the 5-branes just like closed fundamental 
strings break and connect D-branes which is the U-dual scenario because
those strings and branes are in the $SL(2,Z)$-multiplet ~\cite{Schwarz950}.
Thus there are ($ Q_{2} Q_{s5}$) D-twobranes and we
have in general a factor of ($4 \prod_{p\neq N}Q_{p}$).
     
Note that the factor ($4 \prod_{p\neq N}Q_{p}$) is basically the only
string theoretical input because from now on we treat the open strings as an
ideal gas. Ideal means noninteracting and we treat the gas with
classical kinetic gas theory. The gas has ($4 \prod_{p\neq N}Q_{p}$)
types of bosons and ($2 \times 4 \prod_{p\neq N}Q_{p}$) types of
fermionic partners due to supersymmetry where we put another factor of 2
for the two spin states of particles with spin $1/2$. The total energy
is $\frac{Q_{N}}{R_{9}} $ and the ``box'' the gas lives in has the volume
$(2 \pi R_{9} )$. Therefore the entropy follows as 
\begin{eqnarray}
S = \sqrt{\frac{\pi }{6} (1+2) (4 \prod_{p\neq N}Q_{p})
(\frac{Q_{N}}{R_{9}} )	(2 \pi R_{9}) } = 2 \pi \sqrt{\prod_{p}Q_{p}  } 
\end{eqnarray} 
which is exactly equation (\ref{SQ}) and which is only valid for large
charges as well. 

~\cite{Johnson96}, who had $d=3+1$ configurations with only three
charges, express the $d(N)$ possible ways of distributing a total of
$N=Q_{N}$ or $Q_{W}$ units to the possible (1,5) strings or fundamental strings
with help of the partition function 
\begin{eqnarray}
\sum \, d(N) q^{N} = (\prod_{n=1}^{\infty } \frac{1+q^n }{1-q^n } ) ^{4Q_{p}Q_{p+4}}\\
\Rightarrow \lim_{N\rightarrow \infty }(dN) = e^{2 \pi \sqrt{Q_{p}Q_{p+4}N}}
\end{eqnarray}
(similar in ~\cite{Johnson96}) and therefore equation (\ref{SQ}) is
valid again.

\subsubsection{Counting of States for the Near Extremal Black Holes}
The D-brane picture and the counting are almost the same as for the
extremal black holes. Therefore I will describe the supergravity side
later.
 
Near extremal black holes have finite temperature and are therefore not
possibly to be described by the BPS states because finite temperature
implies Hawking radiation but BPS states are stable. Recall that we
added only left moving momentum because adding left and right moving
momentum would destroy the BPS properties. Hence, adding right moving
momentum along $S^{1}_{9}$ should model a non extremal black hole. In
general, adding anti charges ($-Q_{p}$) gives the configuration more
mass than the lowest bound that is required to carry the total charge.
Therefore we have a non extremal configuration and charges may
annihilate anti charges which will result in Hawking radiation. This
works only fine in the near extremal region because then there are so
few right movers for example that one can neglect the interactions
between left moving and the right moving gas. This is called the
``dilute gas approximation'' and it allows us to just add the
entropies of the gases. However, one should recall that the simple
formula for the entropy was deduced for large charges only. In order for
these equations to be valid again we have to add not few but many anti
charges. Thus we require for the left and right moving momentum for
example: $N_{L} \gg N_{R} \gg 1$ where $Q_{N} = N_{L} - N_{R}$. Adding 
the entropies yields \footnote{Because of the discussion leading to 
equation (\ref{Reff}) we could add far less momentum, namely as little
as two units of momentum $1/R_{eff}$ such that \\ 
$S = 2 \pi (\prod_{p\neq N} \sqrt{Q_{p}} )
(\sqrt{N_{L} + \frac{1}{R_{eff}}} +
 \sqrt{\frac{1}{R_{eff}}} )$ and $P_{9}$ of the whole
configuration is still quantised to units of $1/R_{9}$.}
\begin{eqnarray}
S = 2 \pi (\prod_{p\neq N} \sqrt{Q_{p}} ) (\sqrt{N_{L}} + \sqrt{N_{R}} )
\end{eqnarray}  
In general, with 
\begin{eqnarray}
Q_{p} = (N_{p} - N_{\bar{p} }) \label{Np}
\end{eqnarray}
the same is valid for all kinds of branes because U-duality interchanges
momentum and branes among each other. That alters coupling constants and
radii of compactifications but it lets the entropy unchanged because of
its topological origin. Thus 
\begin{eqnarray}
S = 2 \pi \prod_{p}  (\sqrt{N_{p}} + \sqrt{N_{\bar{p} }} ) \label{SNN}
\end{eqnarray}
where $p=1, 5, N$ or $p=2, 6, s5, N$ with $N_{N} = N_{L}$ and
$N_{\bar{N} } = N_{R}$. Equation (\ref{SNN}) can be written as an expansion
that gives the increase due to deviations from the extremal case
precisely: 
\begin{eqnarray}
S = 	2 \pi (\prod_{p} \sqrt{N_{p}} )
	 (1+ \sum_{p} \sqrt{\frac{N_{\bar{p} }}{N_{p} } } + \cdots ) \\
\simeq 	2 \pi (\prod_{p} \sqrt{Q_{p}} )
	 (1+ \sum_{p} \sqrt{\frac{N_{\bar{p} }}{N_{p} } } + \cdots )
\end{eqnarray}
The latter part of this formula gives the leading contributions to the increase in entropy
as being $ \propto \sqrt{\frac{N_{\bar{p} } }{N_{p} } } $ if we add equal
amounts  of charges and anti charges in order to keep the total charge
fixed.

\newpage
\subsubsection{The Supergravity of Near-Extremal d=4+1 and d=3+1 Black Holes}
On the supergravity side one has to find the metric for the general non
extremal black hole. This is done by starting with the extremal
Reissner-Nordstr\"{o}m solution which had the constraints (\ref{cpQp}).
The constraints are removed by boosting the solution in order to change
the momentum and by using U-duality to interchange the charges amongst
each other so that every charge can become a momentum in a dual
description and can then be boosted ~\cite{HoroStrom96} ~\cite{HoroMalda96}.  
This boosting with boost parameters $\alpha _{p} $ results in $\sinh
\alpha _{p}$-type functions familiar from special relativity and the
resulting metric is (\ref{tausumperpar})  with (\ref{unbalanced}) [or 
(\ref{balanced}) for IIA on $T^6 $], (\ref{dependence}) and
(\ref{simplesubstitution}) but instead of 
$ +(1-f_{N}) (dx_{0} - dx_{9})^2 $ in (\ref{simplesubstitution}) write 
\begin{eqnarray}
- \frac{r_{0}^{d-3}}{r^{d-3}} 
	[\cosh \alpha _{N} \, (dx_{0})  + \sinh \alpha _{N} \, (dx_{9}) ]^2 
\end{eqnarray}
and we substitute\footnote{Especially in this section the formulae can
nowhere be found like this and may be hard to compare with the
literature. This notation allows one to be most general and brief.}:
\begin{eqnarray}
(c_{p}^{(d)} Q_{p}) \rightarrow 
	(r^{d-3}_{0} \sinh^{2} \alpha _{p} ) =: r_{p}^{d-3} \label{boostedcpQp}
\end{eqnarray}
The charges are (with $\alpha ' = 1 $ for simplification)
\begin{eqnarray}
 Q_{p} =  (\frac{C_{(d)} }{M_{p}}) \sinh 2\alpha _{p} \label{Qp}
\end{eqnarray}
where  
\begin{eqnarray}
 C_{d} =  (\frac{r_{0}^{d-3} V_{T^{D-d}} }{g^2 2^{d-4}}) \label{Cd}
\end{eqnarray}
with the mass $M_{p}$ again dependent on the tension formulae
[(\ref{Tfund}) etc.]:
\begin{eqnarray}
M_{1} = \frac{R_{9}}{g} & M_{5} = \frac{V_{T^{5}}}{g} & M_{N} 
	=  \frac{1}{R_{9}} \nonumber \\
M_{2} = \frac{R_{9} R_{8}}{g} & M_{6} = \frac{V_{T^{6}}}{g} & M_{s5}
 	= \frac{R_{9}R_{7}R_{6}R_{5}R_{4} }{g^2} 
\end{eqnarray}
Note that the term $\frac{C_{(d)}}{M_{p}} $ has the units of a number
and will therefore appear in equations that are dependent on pure
numbers (e.g.:(\ref{Qp}) and (\ref{Scosh})).

Reduction to $d$ dimensions leads again to equations like (\ref{tauF})
with (\ref{F}) but this time we write: 
\begin{eqnarray}
(d\tau )_{E}^2 = (F^{-1})^{a_{(d)}} [B (dx_{0})^2 ] 
		- F [B^{-1} (dr)^2 + r^2 \, (d\Omega _{(d-2)} )^{2}]
\end{eqnarray}
with $B=(1-\frac{r_{0}^{d-3}}{r^{d-3}}) $ and $ (d\Omega _{(2)} )^{2} =
(d\Theta )^2 + \sin ^2 \Theta \, (d\phi )^2 $ as usual and $F$ from
(\ref{F}) with the boosted $(c_{p}^{(d)} Q_{p})$ from equation
(\ref{boostedcpQp}).
The solution is invariant under interchange of the boost parameters
$\alpha _{p}$ due to the applied U-duality. $r=0$ is still an event
horizon. In the extremal case it was the inner and the outer one since
they coincide. Here it is the inner event horizon and $r=r_{0}$ is the
outer one. Applying the methods of general relativity we find the ADM
mass 
\begin{eqnarray}
 M = C_{(d)} \sum_{p} \cosh 2\alpha _{p} \label{MCcosh}
\end{eqnarray}
and the entropy and temperature
\begin{eqnarray}
S =	\frac{A_{+}^{(d)}}{4G_{N}^{(d)}} = 
	2 \pi \prod_{p} \, [\sqrt{2 \frac{C_{(d)}}{M_{p}}} \, 
	\cosh \alpha _{p}] \label{Scosh} \\
\Rightarrow T_{H} \propto \frac{1}{\prod_{p} \cosh \alpha _{p}} 
\label{Tcosh}
\end{eqnarray}
The compactification leads to $(d-2)$ scalars due to the components
$G_{99}$, $G_{55} = G_{66} = G_{77} = G_{88}$  [or 
$G_{99}$, $G_{88}$, $G_{44} = G_{55} = G_{66} = G_{77}$ in the case of
IIA on $T^{6}$. These are related to the
pressures $P_{\mu }$ along compactified directions as discussed when we
balanced the scalars. It holds simply $P \propto C_{(d)} \times
[$factors of $ \cosh 2 \alpha _{p}]$, for example
\begin{eqnarray}  
P_{5} = P_{6} = P_{7} = P_{8} 
	= C_{(5)} (\cosh 2 \alpha _{1} - \cosh 2 \alpha _{5})
\end{eqnarray}
for IIB on $T^5 $ or for IIA on $T^6 $ for example
\begin{eqnarray}
P_{4} = P_{5} = P_{6} = P_{7}
        = C_{(4)} (\cosh 2 \alpha _{2} - \cosh 2 \alpha _{6})    
\end{eqnarray}
according to which directions the certain D-branes are
parallel or orthogonal to. The extremal black hole has $r_{0} = 0$ and
one or more $\alpha _{p} = \pm \infty $ and if all charges are finite
$(Q_{p} \neq 0)$ we will get back the formulae $T_{H} = 0$ and $S=2\pi
\prod_{p} \sqrt{Q_{p}}$ and the mass $M$ will be the sum of the masses of
each contribution:
\begin{eqnarray}
M_{extr} = (\frac{V_{T^{D-d}}}{g^2 }) \, \sum_{p} r_{p}^{d-3} 
\end{eqnarray}
For only one single charge being finite $(Q_{p} \neq 0)$  and for  
the others $Q_{q \neq p} = 0$ one obtains the line elements for a
single brane or single anti brane or a single unit of momentum. Simple
relations between masses and pressures are obtained ~\cite{HoroLowe96}
(but I write a very different notation); for example for a single
D-p-brane or anti D-p-brane with the mass $M_{p}$ for the case IIB on
$T^5 $: 
$M_{1} = P_{5}$, $M_{5} = -P_{5}$ and $M_{N}$ has $P_{5} = 0$ of course
because the D-1-brane and D-5-brane are orthogonal and parallel
to $S_{5}^{1}$ respectively and the momentum does not go along
$S_{5}^{1}$. 

Why do I write all these equations ? They show that masses, pressures
and charges simply decompose into the contributions of single charge
carriers. We need this insight in order to \emph{derive} the expression
for the numbers of charges $(N_{p})$ on the supergravity side  
although comparing equations (\ref{Qp}), (\ref{MCcosh}) and (\ref{Scosh})
with equations (\ref{Np})(\ref{SNN}) makes a guess like 
$N_{p} \propto e^{2\alpha _{p}} $ obvious. With the equations above we
do not have to guess because they allow to express the solution in terms
of numbers of branes and anti branes and units of left and right moving 
momentum instead of using the boost parameters $\alpha _{p}$ and $r_{0}$,
 $R_{9}$ and $V_{T^{4}} = (R_{8}R_{7}R_{6}R_{5}) $ 
[or boost parameters $\alpha _{p}$ and $r_{0}$, $R_{9}$, $R_{8}$
 and $V_{T^{4}} = (R_{7}R_{6}R_{5}R_{4}) $ in case of IIA on $T^6 $].
The following definitions are equivalent to the variables used in
equation (\ref{Np}):
\begin{eqnarray}
N_{p} 		= \frac{C_{(d)}}{M_{p}} \, \frac{e^{+2\alpha _{p}}}{2} \\
N_{\bar{p} } 	= \frac{C_{(d)}}{M_{p}} \, \frac{e^{-2\alpha _{p}}}{2} 
\end{eqnarray}
Now we can rewrite equations like (\ref{Qp}), (\ref{MCcosh}) etc. in a
simple form, for instance
\begin{eqnarray}
M = \sum_{p} M_{p} (N_{p} + N_{\bar{p} })    
\end{eqnarray}
and equations (\ref{SNN}) and the Bekenstein-Hawking entropy (\ref{Scosh})
are equivalent.\\

Just like  we express the numbers with help of boost parameters,
compactification radii and so on we can express these parameters in
terms of the numbers of branes (for a list see ~\cite{Malda96}). If we
maximise $S$ holding the charges $Q_{p}$ and mass $M$ of the
configuration fixed the compactification parameters and in turn the
expression for the numbers $N_{p}$ will result. Therefore the
configuration is a thermodynamic equilibrium of the contributing
charge carriers; but only in this sense because the right and left
moving gasses have different temperatures and are thought to be 
noninteracting. The different temperatures come from the form of
equation (\ref{Tcosh}) which is written inversely as follows:
\begin{eqnarray}
\frac{1}{T_{H}} \propto \prod_{p}\cosh \alpha _{p} \propto
(\prod_{p \neq q}\cosh \alpha _{p}) (\sqrt{N_{q}} + \sqrt{N_{\bar{q}}})
\end{eqnarray}
for example 
\begin{eqnarray} 
\frac{1}{T_{H}} = \frac{1}{2} (\frac{1}{T_{L}} + \frac{1}{T_{R}} ) 
\end{eqnarray}
From equation (\ref{SNN}) follows $S = S_{L} + S_{R}$ 
\begin{eqnarray} 
\Rightarrow \frac{A_{+}}{4G_{N}} = S_{L} + S_{R} \label{SASLSR}
\end{eqnarray}
For an extremal black hole we have $A_{+} = A_{-}$ and $S_{R} = 0$ and
deviating from extremality the inner and outer horizons separate. This
suggests to generalise equation (\ref{SASLSR}) to
\begin{eqnarray} 
\frac{A_{\pm}}{4G_{N}} = |S_{L} \pm S_{R}| \label{SASLSRabsolute} 
\end{eqnarray}
which actually can be verified in just the way as it was done for
$A_{+}$ alone.  Equation (\ref{SASLSRabsolute}) leads to the surprising form
\begin{eqnarray} 
S_{L/R} = \frac{1}{2} (\frac{A_{+}}{4G_{N}} \pm \frac{A_{-}}{4G_{N}} )\\
; \, \, S_{L} > S_{R}\\
S_{R/L} = \frac{1}{2} (\frac{A_{+}}{4G_{N}} \pm \frac{A_{-}}{4G_{N}} )\\
; \, \, S_{R} > S_{L}
\end{eqnarray}
Both horizons seem to have their own thermodynamic
variables which is a suggestion due to ~\cite{Cvetic96}.  
All this is derived in the dilute gas approximation but looks very
fundamental. 

\subsection{Less special Black Holes}
The results reviewed thus far are very special black holes in that the
compactifications and wrappings (5-brane around $T^{5}$ , one-brane
around $S^{1}$ etc.) imply that the D-brane configurations are quite   
artificial ones. All branes intersect with an angle of either zero or a
half $\pi $. As well, we did not treat rotating solutions any closer.
These more general solutions can be obtained with the
solution generating techniques as there are U-duality including
T-duality at an angle and boosting along all sorts of directions. On
rotating black holes see for example ~\cite{Breckenridge962}  
~\cite{Breckenridge963} ~\cite{Cvetic96}.
On solutions with new angles of intersections see for example
 ~\cite{Behrndt97} ~\cite{Breckenridge97} ~\cite{Costa97}.
The supergravity calculations are complicated because of the many boost
parameters and angles but the results for the entropy and the
temperature
for example match the results from the counting of states of the 
configuration of branes and the results show the topological
character of the degeneracy. Solutions that have branes
intersecting at an arbitrary angle have the entropy depending on this  
\emph{continuous} parameter but still the results match the counting of 
states because these solutions interpolate between the different       
configurations of branes with angles of zero and $\pi /2$. The angle is
the mixing parameter. ~\cite{Costa97} for example found a black hole
that interpolates between a configuration of (D-1-branes plus D-5-branes) 
and a configuration of (3-branes plus 3-branes).
The mixing comes from the fact that T-duality orthogonal to a D-p-brane 
creates a D-(p+1)-brane and T-duality parallel to a D-p-brane reverses
the (p-1) to p process and leads to a D-(p-1)-brane. 
T-duality at an angle will therefore mix the D-(p-1)-brane and the 
D-(p+1)-brane.

An exhaustive
enumeration of all the work done on all kinds of black holes and their
thermodynamics in all kinds of dimensions and sorts of compactifications
would be a very long list indeed. Therefore a few examples out of many  
produced by a whole  industry of finding strange black hole solutions   
must suffice. Other kinds of charges have been considered; for example
black holes with D-0- and D-6-brane charge ~\cite{Shein97}
~\cite{Pierre97}, or D-0- and D-4-branes ~\cite{Johnson96}. Other
dimensions have been treated; for example $d=2+1$ ~\cite{Birm97} 
~\cite{Taejin97} and  $d=1+1$ ~\cite{Gegenberg97}.

\subsection{Hawking Decay}
Charges and anti charges can annihilate each other. For example two
D-strings combine with a D-$\bar{1}$-brane 
$( \uparrow + \uparrow + \downarrow )$ resulting in one D-string with
left and right moving momentum. Open strings with $n_{R}$ units of
momentum $1/R_{eff}$ can combine with left moving strings carrying
$n_{L} = n_{R}$ units of momentum resulting in closed strings with
$P_{9} = 0$  and total mass $M=\frac{2n_{L}}{R_{eff}}$ which in turn can
leave the D-branes and tunnel through the outer horizon in case of a
black hole. The emission of
charged particles ($P_{9} \neq 0$) is suppressed because their momentum
has to be quantised in units of $1/R_{9}$ which needs the right numbers
$n_{L}$ and $n_{R}$ of momentum units that happen to combine and results
in a quite massive particle $(1/R_{9})$ anyway. On emission of charged
particles see ~\cite{Gubser968}. The decay rate 
$d\Gamma _{D} $ is dependent on the mass $k_{0}$ of the emitted particle and
the interaction Hamiltonian $H_{int}$ has to be calculated with the
string amplitude for the joining of open strings. This amplitude $A$ is
calculated ``on the disc'' as is familiar from bosonic string theory.
If the (say) two colliding open strings are a bosonic one and a
fermionic one the emitted particle will be a fermion. Then the
calculation of the D-brane decay to lowest order in string coupling
means calculation of the disc amplitude where one bosonic and one fermionic
Vertex operator are on the boundary of the disc and a fermionic closed
string vertex operator is put into the bulk of the disc. On fermionic
decay see ~\cite{Gubser976} and ~\cite{Das977}.
$g$ is the string \emph{coupling} constant, thus $A\propto g$, and in
order to have $A_{(k_{0} = 0)} = 0$ and $A_{(k_{0})} = A_{(-k_{0})}$ 
we require $A \propto k_{0}^{2}$ ~\cite{Hash96}.      
Denoting the emission process for the configuration of branes 
as $|i> \rightarrow |f>$ one may write:
\begin{eqnarray}
d\Gamma _{D} \propto \frac{d^{d-1} k}{k_{0}} \, 
\delta _{[k_{0}-\frac{2n_{L}}{R_{eff}}]} |<f|H_{int}|i>|^2
\end{eqnarray}
If the initial state is not known an averaging over initial states will
be required. This is the very step that leads to the thermal character
of the radiation because the averaging over the initial states
introduces the occupation number of the left and right moving
oscillators as $\rho _{L/R}$ of $(\frac{k_{0}}{2T_{L/R}})$ as given in
equation (\ref{rho}) ~\cite{Callan96}~\cite{Dhar96}~\cite{Das96}.
The summing over the final states leads simply to the product of left
and right occupation numbers. Averaging and summing leads to 
\begin{eqnarray}
|<f|H_{int}|i>|^2 \rightarrow \frac{R_{eff}^{2}}{n^{2}_{L(k)}}
\sum_{i,f}|<f|H_{int}|i>|^2
\end{eqnarray}
as well and this results with $T_{L} \gg T_{R} \Rightarrow T = 2T_{R}$
in
\begin{eqnarray}
d\Gamma _{D} \propto A_{+} \, \, \rho_{(\frac{k_{0}}{2T_{R}})} \, d^4 k
\end{eqnarray}
~\cite{Das96}.

Due to time symmetry, the formulae hold for absorption and emission but
are \emph{derived} only for low energy radiation because we still need the
limit of large but near extremal configurations for the simplifications 
made. Thus, the Compton wavelength of the radiation is far bigger than
the Schwarzschild radius. 
The condition $T_{L} \gg T_{R}$ can be avoided but the calculations are
involved and beyond the scope of the present work
(~\cite{Dhar96} ~\cite{Das96} \\
~\cite{Das967} ~\cite{MaldaStrom969}). In $d=3+1$ dimensions
for near extremal black
holes and in the dilute gas approximation (although $T_{L} \gg T_{R}$ is
not necessary) holds for the decay rate of the string and D-brane
configurations into S-wave scalar particles (compare equation
(\ref{GammaHawking}): 
\begin{eqnarray}
d\Gamma _{H} = d\Gamma _{D} =  (g_{eff} k_{0}) \,
\, \rho_{(\frac{k_{0}}{2T_{L}})}
\rho_{(\frac{k_{0}}{2T_{R}})} \, d^4 k
\end{eqnarray}
because ~\cite{MaldaStrom969}
\begin{eqnarray}
\sigma _{abs} =  (g_{eff} k_{0}) \, \, \frac{\rho_{(\frac{k_{0}}{2T_{L}})}
\rho_{(\frac{k_{0}}{2T_{R}})}}{\rho_{(\frac{k_{0}}{T_{H}})}}
\end{eqnarray}
where $g_{eff}$ is an effective coupling between left and right moving
oscillators.
Thus, the observer at spatial infinity cannot distinguish between the
semi classical black hole with its grey body radiation and the
configurations of strings and D-branes. Both look the same even if
observed in the high energy radiation.

The configuration of the branes is a bound system and so the vibrations
of the strings are only inside the D-5-branes or D-6-branes for IIB and
IIA respectively. Therefore not all closed strings in 10 dimensions 
can be absorbed ~\cite{Das96}. Only the ones that are scalars in the
uncompactified spacetime are not suppressed. This fits to the 
semi-classical result that at low energies scalars are attracted by the black
hole but vector particles and gravitons are repelled by a so called
``centrifugal barrier''. The grey body spectrum of rotating black 
holes has been investigated by  ~\cite{Gubser969} ~\cite{Cvetic975} 
~\cite{Youm976}. 

\subsection{Domains of Applicability}
In order to ensure validity of the statistical approximations the
charges $Q_{p}$ have to be large ($Q_{p} \gg 1$). On the other hand, in
order to avoid large open string loop corrections to the perturbative
brane picture (Hawking decay, scattering) we would like $gQ_{p} \ll 1$.
Both conditions are possible simultaneously in the limits $g\rightarrow
0$, $Q_{p} \rightarrow \infty $ but $gQ_{p} = $ const. However, having
$G_{N} = 1 = \alpha '$ for simplification and the charges $Q_{p}$
comparable to each other, the metrics like (\ref{tauF}) for example
imply that for the D-branes for example we need $gQ\gg 1$ ($g^2 Q_{N}$
for momentum, the power of g depends on the power of g in the mass
formulae) in order to model a black hole, i.e. in order to have the
Schwarzschild radius larger than the string length 
$\sqrt{\alpha '} = 1$. No BPS properties extend the range of
validity of the D-brane picture in the non extremal cases. Thus, at 
about $gQ \simeq $ string length $\simeq $ gravitational radius 
($\simeq 1$), which is called the correspondence point
~\cite{HoroPolch96}, we are in between the domain of validity of the
D-brane description ($gQ \ll 1$) and the black hole region ($gQ \gg 1$).      
The domain of validity of the models encountered seems to be very
large though because not only does the statics (entropies and
temperatures) of the semi-classical theory and M-theory agree but we
found agreement of the dynamics as well (decay, scattering). This
implies that the open string loop corrections are somehow suppressed (by
a background of open strings suspect ~\cite{Callan96}) and that the
near BPS D-brane picture --- however it actually looks like at 
$gQ \gg 1$ --- is valid in the near extremal black hole region as well.   
Ranges of applicability and where the D-brane description 
and the semi-classical one agree and disagree are further discussed in
~\cite{Haw97} ~\cite{Dowker97} ~\cite{Larsen972}. In fact, closer
investigation reveals that there are phase transitions to be considered
if we want to describe non extremal black holes that are not near
extremal. ~\cite{Mathur976} for example varied 
compactification moduli so that the string theoretical description
breaks down when the curvature comes close to the string length scale.
He found that it becomes entropically advantageous for the configuration
of branes and strings to rearrange when the correspondence point at 
$gQ \simeq 1 $ is crossed. What has been vibrations in the $gQ \ll 1$
region can be turned into solitonic 5-branes for example. This is a
phase transition because the degrees of freedom change. That this
transition is negligible for near extremal black holes 
shows only that the
degeneracy does not change much for those near stable configurations
during the phase transition
although the question of how the configurations look like at strong
coupling might have a surprising answer for all black holes --- whether
non extremal or extremal.   

\subsection{Remarks on Storage and Evolution of Information}
The discussion of whether the information is stored at the singularity
or on
the event horizon is not resolved yet because one is not sure how the  
weakly coupled D-brane configurations look like in the strong coupling
region where they model black holes. This issue lies at the heart of the
information loss problem. The sources of the RR charges are at the
singularity and the D-branes are the carriers of the RR charges. Thus
the strong coupling limit of the D-brane arrangement, which is expected
to be not larger than the string length, should be at the singularity.
But on
the other hand, the thermo\emph{dynamics} of black holes implies that the
information is stored near the horizon because there the black hole
interacts with external probes. Moreover, the grey body factors are due
to a filtering outside the black hole, so the information should be
accessible there as well as if the black holes store the information in
long hair on the event horizon.
If the information is near the singularity, how does it get to the
surface ?
This is addressed by  ~\cite{Hotta97} as the problem of ``... the
mechanism of carrying out the information to the outside...'' and he
suggests a very mechanical mechanism indeed (see below).

Susskind and others think of the information as being stored  in a shell
between the event horizon and a stretched horizon about one string
length further away \cite{SUSK931} ~\cite{SUSK932} ~\cite{SUSK941}
~\cite{SUSK942} \\ ~\cite{SUSK943}.
The stretched horizon was defined as \cite{SUSK9394}  the location where
the local Unruh temperature (formula (\ref{Unruh})) equals the Hagedorn
temperature which is the temperature above which the energy does not
go into more and more excitations of the strings (temperature is not  
increased anymore) but leads to longer and longer strings (a phase    
transition (!)). \cite{Sen95} showed that this is equivalent to the  
definition of the stretched horizon as the location where the theory of
the string world sheet becomes strongly coupled. Susskind put forward the idea of a
``holographic'' world, meaning that the 3+1 dimensional world has so few
degrees of freedom that a 2+1 dimensional description on its 2+1
dimensional boundary is sufficient to account for everything
~\cite{SUSK95}. His argument uses an idea due to t'Hooft
~\cite{Hooft93} based on Bekenstein's observation that the maximum
entropy of a region of space is realised by a black hole which suggests
a mapping of its microstates onto the 2 dimensional event horizon
because the entropy and the area of the horizon are proportional.  I 
recommend Susskind's work here because he gives an argument
involving black holes and their microstates due to string theory that is
moreover in parts very accessible even for people without much knowledge
of the related physics ~\cite{SUSK95}.

`t Hooft took the Planck instead of the string length and derived the
thermodynamics with help of scalar fields living outside the black hole
~\cite{Hooft85}. t'Hooft proposed an S-matrix approach which is
necessarily unitary.
The first S-matrix ansatz showed that the gravitational interaction between
ingoing particles and outgoing Hawking radiation can act as a mechanism to
recover the information. The effect is a shock wave and outgoing
trajectories are shifted ~\cite{Dray85} ~\cite{Hooft96}.
Hawking did not take into account that the particles of the 
in- and outgoing radiations alter the metric. Particles that fall to
the event horizon have very high energy and therefore the interactions
with outgoing particles should not be neglected. Coming from S-matrix
considerations in gravity ~\cite{Haro97} rediscovered the action of
bosonic string theory. Hence all these suggestions rather complete than
contradict each other.

\cite{Hotta97} describes the gravitational collapse as a phase
transition starting at
the centre of the collapsing body where the degeneracy pressure of any
entropy carrying mass is overcome first. A body of so called ``Planck
solid'' develops pushing the mass (and the information with it) in form
of a layer
of string gas outside the Schwarzschild radius. This gas is then between
stretched and Planck horizons and leads to Hawking radiation with the
inside of the black hole being smooth.

\cite{Horo96} describe the information as being carried by modes that 
live only inside the event horizon but that stretch between it and the
singularity and do not travel into other asymptotically flat spacetimes
beyond the singularity. There are a large number of such modes that
leave the event horizon smooth. It is shown how these modes are
generated at the singularity where the RR charges sit. These interior
modes are inhomogeneous in the internal (compactified) dimensions and
the whole picture of these modes can be thought of as being like the 
field of a shell of charge in electrostatics. The field is trivial   
inside but non-trivial outside. Here one finds an ``inside-out'' line  
element, i.e. with increasing $r$ one gets closer to the singularity,
and for these modes the fields outside instead of the inside are
trivial. The modes would have to be faster than light in order to
carry information from the singularity to the
event horizon. This is overcome by the topology having closed timelike
loops (these are loops that are timelike along every section, so time
repeats itself over and over for bodies with such worldlines).

Black hole complementarity allows us to have two different descriptions
simultaneously. One of them may carry the information on the event horizon and
the other one inside the black hole. The D-brane picture shows a possible
origin of this complementarity. Everything that falls onto the event
horizon is turned into open strings on the branes which are ---
in this description --- at the event horizon at say $r=0$ where we have
put the carriers of the charges. In order to compare the classical
picture with the D-brane picture one has to transform the coordinates
because the system with the $r$ from above does not cover the inside of
the black hole (see the discussion after equation (\ref{F})).
By falling inside the black hole the closed strings are turned into open
ones on the branes and the ongoings inside the black hole are actually
interactions of the open strings on the branes. No information will be
lost. At first it seems
strange that for example human observers and their measurement devices
should not realise that they hit the branes and are turned into open 
strings because the equivalence principle says that they feel nothing
unusual but the approaching singularity. On
the other hand, in the model of the holographic universe, the observer
has always been 2+1 dimensional. The D-brane picture might be a crucial
step in finding the mapping between the 3+1 dimensional description and
the 2+1 dimensional holographic one.   

The maybe most radical position is taken by ~\cite{Amati97} who claims
that not only the thermal spectrum but the black hole itself, its event
horizon and the whole causal spacetime description are only due to an  
averaging over initial states that basically equals a classical
treatment. Quantum mechanically the black hole does not exist as such.
This reminds of (M)atrix and twistor theories which treat spacetime as
an emergent classical limit. With the discussion of section (\ref{0-3})
in mind one might like to ask whether the event horizons exist only
with a ``overwhelming probability for systems with a large number of
degrees of freedom''.

\newpage
\part{}
\section{Conclusions}
We have seen that the semi-classical descriptions and the stringy ones
of black holes and their thermodynamics lead to very similar results.
The string theoretical descriptions are the ones that can account for the
microstates which are topologically and geometrically well understood
(at least at weak coupling). 
The agreement in statics and dynamics for the extremal and near
extremal regions is striking and
this suggests the fundamental nature and validity of some findings for 
all black holes. For example the two horizons both seem to have their
own thermodynamics. A general non extremal and not near extremal black
hole can not yet be described with M-theory because the string
theoretical approach might not be sufficient or at least has to be
altered considerably due to phase transitions. The underlying
superconformal field theory is not yet found in detail. 

Although there are strong hints in favour for the unitarity of black
hole evolution  it is
too early to decide in favour for any one of the many models and it will
be too early as long as there is no agreement over the exact form of the
entropy of a say small sized, simple Schwarzschild black hole and no
understanding of what its microstates are. As long as we have good
results only near extremal, all one knows is that the theories work so
well in that region because the shortcommings they have are negligible there.
What we may expect is that both --- semi-classical and superstring theory  
--- will go on and try to reproduce each others results until the theory
that is valid even for a simple and small black hole is found. This
theory will be a major step forward for all those reasons for which the
simple black holes are so famous --- like the question of how many 
degrees of freedom
has the universe --- and because one expects a consistent theory of
quantum gravity. 

Black holes are likely to go the way the hydrogen atom did: From 
``oh-so-strange
and fundamental'' to just another ``chemicists thing''. Probably, the
authors of popular science, who could make capital out of concentrating
on the strange things that seem to happen when one takes the theory of
general relativity at the points where serious scientists only
concentrate because they expect it to break down, will see all those
other universes and time machines disappear. Then we are left with grey
bodies called ``black'' holes because the gravitation is at some distance
strong enough to accelerate light inwards -- oh well, so what ? 

\section{Acknowledgements}
I would like to thank my supervisor Professor David Bailin for his guidance
through my MSc course. I am grateful to him for many helpful
discussions. I am grateful to Doctor Mark B. Hindmarsh who had always
time for my questions and who helped me with the computers.
As well there is to mention the support provided by the physics
departement of the University of Sussex allowing me to attend the String
'97 conference where I had many useful conversations that contributed to
this MSc-thesis.
 
\section{General Bibliography and References}
To be recommended as texts on general relativity:
\begin{itemize}
\item Foster J. and Nightingale J.D. \emph{A short course in General
	Relativity}, Springer Verlag, New York 1995
\item Hawking S.W. and Ellis G.F.R. \emph{The large scale structure of the Universe},
	Cambridge University Press, Cambridge 1973
\item Wald R.M. \emph{General Relativity}, University of Chicago Press,
        Chicago 1984
\item Wald R.M. \emph{Quantum Field Theory in Curved Space-Time and
	Black Hole Thermodynamics}, University of Chicago Press,
        Chicago 1994
\end{itemize}
Texts for the introduction into superstring theory:
\begin{itemize}
\item Bailin D. and Love A. \emph{Supersymmetric Gauge Theory and String
	Theory}, Grad. Student Series in Physics, IOP Publishing, Bristol 1994
\item Green M., Schwarz J. and Witten E. \emph{Superstring Theory},
	Cambridge Univ. Press, 1987
\item Kaku M. \emph{Introduction to Superstrings}, Grad. texts in
        Contemporary Phys., Springer Verlag,New York 1988
\item Kaku M. \emph{Strings, Conformal Fields, and Topology: An Introduction},
 	 Springer Verlag, 1991
\item Rosenbaum \emph{Relativity,Supersymmetry and Strings},
	A., Plenum Press, New York 1990
\end{itemize}
{999}

\begin{thebibliography}{longlonglongerevenlabel}
\bibitem[Amati,97]{Amati97}Amati D., 1997
 \emph{A string piloted understanding of black hole loss of
 quantum coherence}, hep-th/9706157
\bibitem[Asht./Schill.,97]{Ash97}Ashtekar A. and Schilling T.A., 1997
 \emph{Geometrical Formulation of Quantum Mechanics}, gr-qc/9706069
\bibitem[Bachas,95]{Bach95}Bachas C., 1995
 \emph{D-brane dynamics}, hep-th/9511043
\bibitem[Banks et al,96]{Banks96}Banks T. et al, 1996
 \emph{M-Theory as a Matrix Model: A Conjecture}, Phys. Rev. {\bf D55}
 5112-5128, hep-th/9610043
\bibitem[Barbon,96]{Barb96}Barbon J.L.F., 1996
 \emph{D-brane form-factors at high energy}, hep-th/9601098
\bibitem[Behrndt/Cvetic,97]{Behrndt97}Behrndt K. and Cvetic M., 1997
 \emph{BPS-Saturated Bound States of Tilted P-Branes in TypeII String Theory},
 hep-th/9702205
\bibitem[Bek.,73/74]{Bekenstein73}Bekenstein J.D., 1973/4
 \emph{}, Phys. Rev. {\bf D7}, 2333 / {\bf D9},3292
\bibitem[Bek.,75]{Bekenstein75}Bekenstein J.D., 1975
 \emph{}, Phys. Rev. {\bf D12},3077
\bibitem[Bimo./Liber.,97]{Belg96}Belgiorno F. and Liberati S., 1996
 \emph{Black Hole Thermodynamics, Casimir Effect and Induced Gravity}, gr-qc/9612024
\bibitem[Bimo. et al,97]{Bimonte97}Bimonte G. et al, 1997
 \emph{2+1 Einstein Gravity as a Deformed Simons Theory}, hep-th/9706190
\bibitem[Birm./Sachs/Sen,97]{Birm97}Birmingham D., Sachs I. and Sen S., 1997
 \emph{Three Dimensional Black Holes and String Theory}, hep-th/9707188
\bibitem[Brans/Dicke,73]{Brans61}Brans C. and Dicke R.H., 1961
 \emph{}, Phys. Rev. {\bf 124}, 925
\bibitem[Brek. et al,96,2]{Breckenridge962}Bekenbridge J.C. et al, 1996
 \emph{D-branes and Spinning Black Holes}, Phys. Lett. {\bf B381},
 423-426, hep-th/9602065
\bibitem[Brek. et al,96,3]{Breckenridge963}Bekenbridge J.C. et al, 1996
 \emph{Macroscopic and Microscopic Entropy of Near-Extremal Spinning
Black Holes},
 hep-th/9603078
\bibitem[Brek. et al,97]{Breckenridge97}Bekenbridge J.C. et al, 1997
 \emph{New angles on D-branes}, hep-th/9703041
\bibitem[Callan et al,91]{Callan91}Callan C., Harvey J. and
 Strominger A., 1993
 \emph{}, Nucl. Phys. {\bf B359}, 611
\bibitem[Callan et al,95]{Callan95}Callan C., Maldacena J.M. and
 Peet A.W., 1995
 \emph{Extremal Black Holes as Fundamental Strings}, Nucl. Phys. {\bf B},
 hep-th/9510134
\bibitem[Callan/Malda.,96]{Callan96}Callan C. and Maldacena J.M.,1996
 \emph{D-brane Approach to Black Hole Quantum Mechanics}, Nucl. Phys.
 {\bf B472}, 591-610, hep-th/9602043
\bibitem[Campa./Lousto,93]{Campanelli93}Campanelli M. and Lousto C.D., 1993
 \emph{}, Int. J. Mod. Phys.{\bf D2},457
\bibitem[Camp./West,84]{Campbell84}Campbell I. and West P., 1984
 \emph{}, Nucl. Phys. {\bf 243}, 112
\bibitem[Carlip,95]{Carlip95}Carlip S., 1995
 \emph{Statistical Mechanics and Black Hole Entropy}, gr-qc/9509024
\bibitem[Carter,71]{Carter71}Carter B., 1971
 \emph{}, Phys. Rev. Lett. {\bf 26}, 331
\bibitem[Chandra.,83]{Chandra.S.83}Chandrasekhar S., 1983
 \emph{The Mathematical Theory of Black Holes}, Oxford,Clarendon Press
\bibitem[Cohen et al,86/87]{Cohen86/87}Cohen A. et al, 1986/87
 \emph{}, Nucl. Phys. {\bf B267},143/{\bf B281},127
\bibitem[Corrigan,75]{Corrigan75}Corrigan E.F. and Fairlie D.B., 1975
 \emph{}, Nucl. Phys. {\bf B91},527
\bibitem[Costa/Cvetic,97]{Costa97}Costa S. and Cvetic M., 1997
 \emph{Non-threshold D-brane bound states and black holes with non-zero
 entropy}, hep-th/9703204
\bibitem[Cvetic/Youm,95]{Cvetic950}Cvetic M. and Youm D., 1995
 \emph{Dyonic BPS saturated  Black Holes of Heterotic String Theory on a
six torus}, Phys. Rev{\bf D53}, 584, hep-th/9507090
\bibitem[Cvetic/Tsey.,95]{Cvetic95}Cvetic M. and Tseytlin A., 1995
 \emph{Solitonic Strings and BPS saturated dyonic Black
 Holes}, Phys. Rev{\bf D53}, 5619-5633, hep-th/9512031
\bibitem[Cvetic/Youm,96]{Cvetic96}Cvetic M. and Youm D., 1996
 \emph{Entropy of Non-Extreme Charged Rotating Black Holes in String 
 Theory}, Phys. Rev. {\bf D54}, 2612-2620, hep-th/9603147
\bibitem[Cvetic/Larsen,97]{Cvetic975}Cvetic M. and Larsen F., 1997
 \emph{General Rotating Black Holes in String Theory: Greybody Factors
 and Event Horizons}, hep-th/9705192, and  \emph{Greybody Factors for
 Rotating Black Holes in Four Dimensions}, hep-th/9706071
\bibitem[Dabhol. et al,89/90]{Dab8990}Daholkar A. and Harvey J., 1989
 , Phys. Rev. Lett. {\bf 63}, 478 and Gibbons G., Harvey J. and
 Ruiz-Ruiz F., 1990, Nucl. Phys. {\bf B340 }, 33
\bibitem[Dai et al,89]{Dai89}Dai J., Leigh R.G., Polchinski J., 1989
 \emph{New Connections Between String Theories}, 
 Mod. Phys. Lett. {\bf A4},2073-2083
\bibitem[Das/Mathur,96,6]{Das96}Das S. and Mathur S., 1996
 \emph{Comparing Decay Rates of Black Holes and D-branes}, Nucl. Phys.
 {\bf B478},561-576, hep-th/9606185
\bibitem[Das/Mathur,96,7]{Das967}Das S. and Mathur S., 1996
 \emph{Interactions involving D-branes}, hep-th/9607149
\bibitem[Das et al,97]{Das977}Das S. et al, 1997
 \emph{Black hole fermionic radiance and D-brane decay}, hep-th/9707124
\bibitem[Dhar et al,96]{Dhar96}Dhar A., Mandal G. and Wadia S.R., 1996
 \emph{Absorbtion versus Decay Rates of Black Holes in String Theory},
 Phys. Lett. {\bf B388}, hep-th/9605234
\bibitem[Dick,97]{Dick97}Dick R., 1997
 \emph{The string scale and the Planck scale}, hep-th/9707195
\bibitem[Dijk. et al,92]{Dijk92}Dijkgraaf R., Verlinde E. and H., 1992
 \emph{}, Nucl. Phys. {\bf B371},269 
\bibitem[Douglas,95]{Douglas95}Douglas M., 1995
 \emph{Branes within Branes}, hep-th/9512077
\bibitem[Douglas et al,96]{Douglas961}Douglas M. et al, 1996
 \emph{D-Branes and short Distances in String Theory}, hep-th/9608024
\bibitem[Douglas,96]{Douglas962}Douglas M., 1996
 \emph{Superstring Dualities, Dirichlet Branes and the Small
 Scale Structure of Space}, hep-th/9610041
\bibitem[Dowker et al,96]{Dowker97}Dowker H.F., Kastor D. and Traschen J., 1997
 \emph{U-duality, D-branes and black hole emission rates: agreements and
 disagreements}, hep-th/9702109
\bibitem[Duff,96]{Duff96}Duff M., 1996
 \emph{M-Theory: The Theory Formerly Known As Strings}, hep-th/9608117
\bibitem[Dray/Hooft,85]{Dray85}Dray T. and t'Hooft G., 1985
 \emph{The Gravitational Shock Wave of a Massless
 Particle}, Nucl. Phys. {\bf B253}, 173-188
\bibitem[Foster/Night.,95]{Foster95}Foster J. and Nightingale J.D., 1995
 \emph{A short course in General Relativity}, Springer Verlag, New York 1995
\bibitem[Garfinke,92]{Garf92}Garfinke D.,1992
 \emph{}, Phys. Rev. {\bf D46}, 4286
\bibitem[Gegen./Kuns.,97]{Gegenberg97}Gegenberg J. and Kunstatter G., 1997
 \emph{Solitons and Black Holes}, hep-th/9707181 
\bibitem[Gibb./Hawk.,77]{Gibb77}Gibbons G.W. and Hawking S.W., 1977
 \emph{}, Phys. Rev. {\bf D15}, 2752
\bibitem[Giveon,91]{Giveon91}Giveon A., 1991
 \emph{}, Mod. Phys. Lett. {\bf A6},2843
\bibitem[Gosh/Mitra,96/97]{Ghosh9697}Ghosh A. and Mitra P., 1996
 \emph{}, Phys. Rev. Lett.{\bf 77}, 4848 and 1997 {\bf 78}, 1858
\bibitem[Green,76]{Green76}Green M.B., 1976
 \emph{},  Nucl. Phys. {\bf B103}, 333
\bibitem[Green et al,82]{Green82}Green M.B. and Schwarz J., 1982
 , Phys. Lett. {\bf B}, 444
\bibitem[Gubser/Kleb.,96,8]{Gubser968}Gubser S.S. and Klebanov I.R., 1996
 \emph{Emission of Charged Particles from Four- and Five- Dimensional
Black Holes}, Nucl. Phys. {\bf B482} 173, hep-th/9608108
\bibitem[Gubser/Kleb.,96,9]{Gubser969}Gubser S.S. and Klebanov I.R., 1996
 \emph{Four Dimensional Greybody Factors and the Effective String}, Phys. 
 Rev. Lett. {\bf 77} 4491, hep-th/9609076
\bibitem[Gubser,97]{Gubser976}Gubser S.S., 1997
 \emph{Absorption of photons and fermions by black holes in four
 dimensions}, hep-th/9706100
\bibitem[Hadley,97]{Hadley97}Hadley M.J., 1997
 \emph{The Logic of Quantum Mechanics Derived from Classical
 General Relativity}, Found. Phys. Lett. {\bf 10}, 43-60,
 quant-ph/9706007
\bibitem[Haro,97]{Haro97}de Haro Olle S., 1997
 \emph{Noncommutative Black Hole Algebra and String Theory from
 Gravity}, hep-th/9707042
\bibitem[Hash./Kleb.,96]{Hash96}Hashimoto A. and Klebanov I., 1996
 \emph{Decay of excited D-branes}, hep-th/9604065
\bibitem[Hawk.,72,1]{Haw72}Hawking S.W., 1972
 \emph{}, Commun. Math. Phys. {\bf 25}, 152
\bibitem[Haw.,72,2]{Hawking72}Hawking S.W., 1972
 \emph{}, Commun. Math. Phys. {\bf 25},167
\bibitem[Hawk.,75]{Hawking75}Hawking S.W., 1975
 \emph{}, Commun. Math. Phys. {\bf 43}, 199-220
\bibitem[Hawk.,75/76]{Hawking7576}Hawking S.W., 1975/6
 \emph{}, Commun. Math. Phys. {\bf 43}, 199-220 and \emph{Breakdown of
 Predictability in Gravitational Collapse} Phys. Rev. {\bf D14}, 2460
\bibitem[Hawk.,76]{Hawking76}Hawking S.W., 1976
 \emph{}, Phys. Rev. {\bf D13}, 191
\bibitem[Hawk. et al,95/96]{HawHor956}Hawking S.W., Horowitz
 G.T. and Ross S.F. 1995
 \emph{}, Phys. Rev. {\bf D51}, 4302
 and Hawking S.W. and  Horowitz G.T.,1996
 \emph{}, Class. Quantum Grav.{\bf 13}, 1487
\bibitem[Hawk./Tayl.,76]{Haw97}Hawking S.W. and Taylor-Robinson M.M., 1997
 \emph{Evolution of near extremal black holes}, Phys. Rev. {\bf D55},
 7680, hep-th/9702045
\bibitem['t Hooft,85]{Hooft85}'t Hooft G., 1985
 \emph{}, Nucl. Phys. {\bf B256}, 727
\bibitem['t Hooft,90/91]{Hooft9091}'t Hooft G., 1990/91
 \emph{}, Nucl. Phys. {\bf B335}, 138/Phys. Scr. {\bf T36}, 247
\bibitem['t Hooft,93]{Hooft93}'t Hooft G., 1993
 \emph{Dimensional Reduction in Quantum Gravity}, gr-qc/9310006
\bibitem['t Hooft,96]{Hooft96}'t Hooft G., 1996
 \emph{The Scattering Matrix
 Approach for the Quantum Black Hole}, gr-qc/9607022
\bibitem[Horava,89]{Horava89}Horava P., 1989
 \emph{}, Phys. Lett. {\bf B231}, 251
\bibitem[Horava/Witten,96]{Horava96}Horava P. and Witten Ed, 1996
 \emph{}, Nucl. Phys. {\bf B460}, 505 and Nucl. Phys. {\bf B475},94
\bibitem[Horo./Strom.,96]{HoroStrom96}Horowitz G.T. and Strominger A., 1996
 \emph{Counting States of Near Extremal Black Holes}, Phys. Rev. Lett.
 {\bf 77}, 2368, hep-th/9602051
\bibitem[Horo./Malda./Strom.,96]{HoroMalda96}Horowitz G.T., Maldacena J. 
 and Strominger A., 1996
 \emph{Nonextremal Black Hole Microstates and U-duality}, Phys. Lett.
 {\bf B383}, 151-159, hep-th/9603109
\bibitem[Horo./Lowe./Malda.,96]{HoroLowe96}Horowitz G.T., Lowe D. 
 and Maldacena J., 1996
 \emph{Statistical Entropy of Nonextremal Four-Dimensional Black Holes
  and U-duality}, hep-th/9603195
\bibitem[Horo./Marolf,96]{Horo96}Horowitz G.T. and Marolf D., 1996
 \emph{Where Is The Information Stored In  Black Holes?}, hep-th/9610171
\bibitem[Horo./Polch.,96]{HoroPolch96}Horowitz G.T. and Polchinski J., 1996
 \emph{A Correspondence Principle for Black Holes and Strings},
 hep-th/9612146
\bibitem[Hotta,97]{Hotta97}Hotta K., 1997 
 \emph{The Information Loss Problem of Black Hole and ...}, hep-th/9705100
\bibitem[Hu,96]{Hu96}Hu B.L., 1996
 \emph{General Relativity as Geometro-Hydrodynamics},  gr-qc/9607070
\bibitem[Hull/Town.,94]{Hull95}Hull C.M. and Townsend P.K., 1994
 \emph{}, Nucl. Phys. {\bf B438}, 109, hep-th/9410167
\bibitem[Hull/Town.,95]{Hull951}Hull C.M. and Townsend P.K., 1995
 \emph{}, Nucl. Phys. {\bf B451}, 525, hep-th/9505073
\bibitem[Israel,67/68]{Israel67}Israel W., 1967/68
 \emph{}, Phys. Rev. {\bf 164}, 1776 and Commun. Math. Phys. {\bf 8}, 245
\bibitem[Jackiw/Teitel.,84]{Jack84}Jackiw R. and Teitelboim C., 1984,
 in \emph{Quantum Theory of Gravity}, edit. by Christensen S.M. and
 Hilger A., Bristol
\bibitem[John. et al,96]{Johnson96}Johnson C.V., Khuri R.R. and Meyers
 R.C., 1996 \emph{Entropy of 4D Extremal Black Holes}, Phys. Lett. {\bf B378},
 78-86, hep-th/9603061
\bibitem[Kallosh,92]{Kallosh92}Kallosh R., 1992
 \emph{}, Phys. Lett. {\bf B282}, 80
\bibitem[Kallosh et al,92]{Kall92}Kallosh R. et al, 1992
 \emph{}, Phys. Rev. {\bf D46}, 5278, hep-th/9205027
\bibitem[Kim,97]{Kim97}Kim H., 1997
 \emph{Thermodynamics of Black Holes in Brans-Dicke Gravity},
 gr--qc/9706044
\bibitem[Kleb./Thor.,95]{Kleb95}Klebanov I.R. and Thorlacius L., 1995
 \emph{The size of p-branes}, hep-th/9510200
\bibitem[Larsen/Wilc.,95]{Larsen95}Larsen F. and Wilczek A., 1995
 \emph{Internal Structure of Black Holes}, Phys. Lett. {\bf
 B375},37-42 (1996), hep-th/9511064
\bibitem[Larsen,97]{Larsen972}Larsen Finn, 1997
 \emph{A String Model of Black Hole Microstates}, Phys. Rev. {\bf D56},
 1005-1008, hep-th/9702153
\bibitem[Leigh,89]{Leigh89}Leigh R.G., 1989
 \emph{Dirac-Born-Infeld Action from Dirichlet $\sigma$-Model},
 Mod. Phys. Lett.{\bf A4}, 2767-27772
\bibitem[Liber./Poll.,97]{Liberati97}Librati S. and Pollifrone G., 1997
 \emph{Entropy and topology for gravitational instantons}, hep-th/9708014
\bibitem[Losev et al,97]{Losev97}Losev A. at al, 1997
 \emph{M and m's}, hep-th/9707250
\bibitem[Malda.,96]{Malda96}Maldacena J.M., 1996
 \emph{Black Holes in String Theory}, hep-th/9607235
\bibitem[Malda.,97]{Malda97}Maldacena J.M., 1997
 \emph{Probing near extremal black holes with D-branes}, hep-th/9705053
\bibitem[Malda./Strom.,96,3]{MaldaStrom96}Maldacena J.M. and Strominger A., 1996
 \emph{Statistical Entropy of Four-Dimensional Extremal Black Holes},
 Phys. Rev. Lett. {\bf 77}, 428-429, hep-th/9603060
\bibitem[Malda./Strom.,96,9]{MaldaStrom969}Maldacena J.M. and Strominger A., 1996
 \emph{Black Hole Greybody Factors and D-Brane Spectroscopy},
 Phys. Rev. {\bf D55}, 861-870 (1997), hep-th/9609026
\bibitem[Mahar./Schwarz,93]{Maharana93}Maharana J. and Schwarz J., 1993
 \emph{}, Nucl. Phys. {\bf B390}, 3
\bibitem[Mashk.,97]{Mash97}Mashkevich V.S.,1997 
 \emph{Conservative Model of Black Hole and Lifting of the
 Information Loss Paradox}, gr-qc/9707055
\bibitem[Mathur,97]{Mathur976}Mathur S.D.,1997 
 \emph{Emission rates, the Correspondence Principle and the Information
 Paradox}, hep-th/9706151 
\bibitem[Nunez et al,96]{Nun96}Nunez D., Quevedo H. and Sudarsky D., 1996
 \emph{}, Phys. Rev. Lett. {\bf 76}, 571
\bibitem[Olive/Witten,78]{Witten782}Olive D. and Witten Ed, 1978	
 \emph{}, Phys. Lett. {\bf B78}, 97
\bibitem[Open Uni.,79]{Open79}
 \emph{Understanding Space and Time, Block 6: Topics in Space
 and Time}, Open Univ. Press, 35-39
\bibitem[Ortiz et al,97]{Ortiz97}Ortiz M.E. and Vendrell F., 1997	
 \emph{Path integrals, black holes and configuration space topology},
 hep-th/9707177
\bibitem[Pierre,97]{Pierre97}Pierre J.M., 1997
 \emph{Comparing D-branes and Black Holes with 0- and 6-brane Charge},
 hep-th/9707102 
\bibitem[Polch.,94]{Polch94}Polchinski J., 1994
 \emph{}, Phys.Rev {\bf D50},6041
\bibitem[Polch.,95]{P95}Polchinski J., 1995
 \emph{Dirichlet-Branes and Ramond-Ramond Charges}, Phys. Rev. Lett.
 {\bf 75}, 4724, hep-th/9510017
\bibitem[Polch. et al,96]{POL96}Polchinski J. Chaudhuri S.
 and Johnson C., 1996
 \emph{Notes on D-Branes}, hep-th/9602052
\bibitem[Polch.,96,7]{Pol961}Polchinski J., 1996
 \emph{String Duality: A Colloquium}, hep-th/9607050
\bibitem[Polch.,96,11]{Pol962}Polchinski J., 1996
 \emph{TASI Lectures On D-Branes}, hep-th/9611050
\bibitem[Rajar.,82]{Raj82}Rajaraman R., 1982
 \emph{Solitons and Instantons}, North-Holland Publishing Company,
 Amsterdam   
\bibitem[Russo/Sussk.,94]{Russo94}Russo J. and Susskind L., 1994
 \emph{Asymptotic level density in heterotic string theory and
 rotating black holes}, hep-th/9405117
\bibitem[Schwarz,93]{Sch93}Schwarz J. H., 1993
 \emph{Does string theory have a duality symmetry relating weak
 and strong coupling?}, hep-th/9307121
\bibitem[Schwarz,95,8]{Schwarz950}Schwarz J. H., 1995
 \emph{An SL(2,Z) Multiplet of Type IIB Superstrings}, hep-th/9508143
\bibitem[Schwarz,95,9]{Sch951}Schwarz J. H., 1995
 \emph{Superstring Dualities}, hep-th/9509148
\bibitem[Schwarz,95,10]{Sch952}Schwarz J. H., 1995
 \emph{The power of M-theory}, hep-th/9510086
\bibitem[Schwarz,96,1]{Sch962}Schwarz J. H., 1996    
 \emph{M theory extensions of T-duality}, hep-th/9601077
\bibitem[Schwarz,96,7]{Sch961}Schwarz J. H., 1996    
 \emph{The Second Superstring Revolution}, hep-th/9607067
\bibitem[Schwarz,96,7,2]{Sch963}Schwarz J. H., 1996
 \emph{Lectures on superstring and M-theory dualities}, hep-th/9607201
\bibitem[Sen,93]{Sen93}Sen A., 1993
 \emph{}, Nucl. Phys. {\bf D404}, 109         
\bibitem[Sen,95]{Sen95}Sen A., 1995
 \emph{Extremal Black Holes and Elementary String
 States}, Mod. Phys. Lett. {\bf A10},2081, hep-th/9504147          
\bibitem[Sen,96]{Sen96}Sen A., 1996
 \emph{Unification of string dualities}, hep-th/9609176          
\bibitem[Shein.,97]{Shein97}Sheinblatt H., 1997
 \emph{Statistical Entropy of an Extremal Black Hole with 0- and 6-Brane
 Charge}, hep-th/9705054
\bibitem[Shenker,95]{Shenker95}Shenker S.H., 1995
 \emph{Another length scale in string theory?}, hep-th/9509132
\bibitem[Starob.,73]{Starob73}Starobinskii A.A., 1973
 \emph{Amplification of Waves during Reflection from a rotating
 Black Hole}, Soviet Phys. -JETP,37,(28-32)
\bibitem[Strom./Vafa,96]{S+V96}Strominger A. and Vafa C., 1996
 \emph{Microscopic Origin of the Bekenstein-Hawking Entropy}, 
 Phys. Lett. {\bf B379}, 99-104, hep-th/9601029
\bibitem[Sussk. et al,93]{SUSK931}Susskind L. et al, 1993
 \emph{The Stretched Horizon and Black Hole Complementarity}, 
 Phys. Rev. {\bf D48}, 3743,hep-th/9306069
\bibitem[Sussk.,93,7]{SUSK932}Susskind L., 1993
 \emph{String Theory and the Principle of Black Hole
 Complementarity}, Phys. Rev. Lett. {\bf 71}, 2367, hep-th/9307168
\bibitem[Sussk.,93,8]{SUSK942}Susskind L., 1993
 \emph{Strings, Black Holes and Lorentz Contraction}, 
  Phys.Rev.{\bf D49}, 6606, hep-th/9308139
\bibitem[Sussk.,93/94]{SUSK9394}Susskind L., 1993/4
 \emph{Some speculations about black hole entropy in string
 theory}, Phys. Rev. {\bf D52}, 6997, hep-th/9309145 and Susskind
 L. and Uglum J., 1994 
 \emph{Black Hole Entropy In Canonical Quantum Gravity And
 Superstring Theory}, Phys. Rev. {\bf D50}, 2700-2711, hep-th/9401070
\bibitem[Sussk./Thor.,94]{SUSK941}Susskind L. and Thorlacius, 1994
 \emph{Gedanken Experiments Involving Black Holes}, Phys.Rev.{\bf D49},966,hep-th/9308100
\bibitem[Sussk./Uglum,94]{SUSK943}Susskind L. and Uglum J., 1994
 \emph{Black Hole Entropy In Canonical Quantum Gravity And
 Superstring Theory}, Phys. Rev. {\bf D50}, 2700-2711, hep-th/9401070
\bibitem[Sussk.,95]{SUSK95}Susskind L., 1995
 \emph{The world as a hologram}, J. Math. Phys.{\bf 36}, 6377
\bibitem[Taejin,97]{Taejin97}Taejin L., 1997
 \emph{(2+1) Dimensional Black Hole and (1+1) Dimensional Quantum
 Gravity}, hep-th/0706174
\bibitem[Thor.,97]{Thorlacius97}Thorlacius L., 1997
 \emph{Introduction to D-branes}, hep-th/9708078
\bibitem[Town.,96]{Town96}Townsend P., 1996
 \emph{Four Lectures on M-Theory}, hep-th/9612121
\bibitem[Town.,97]{TOWN97}Townsend P., 1997
 \emph{(M)embrane Theory on $T^{9}$ }, hep-th/9708034
\bibitem[Tsey.,96]{Tseylin96}Tseytlin A., 1996
 \emph{Extreme dyonic black holes in string theory}, hep-th/9601177 
\bibitem[Unruh,76]{Unruh76}Unruh W.G., 1976
 \emph{}, Phys. Rev. {\bf D14}
\bibitem[Vafa,95,11]{Vafa9511}Vafa C., 1995
 \emph{Gas of D-branes and Hagedorn density of BPS-states}, hep-th/9511088 
\bibitem[Vafa,95,12]{Vafa9512}Vafa C., 1995
 \emph{Instantons on D-branes}, hep-th/9512078 
\bibitem[Vafa,96]{Vafa96}Vafa C., 1996
 \emph{Evidence for F-theory}, hep-th/9602022 
\bibitem[Wald,72]{R.Wald.72}Wald R.M., 1972
 \emph{Electromagnetic Fields and Massive Bodies}, Phys.Rev.D6 1476-1479    
\bibitem[Wald,74]{Wald74}Wald R.M., 1974 
 \emph{Gedanken Experiment to Destroy a Black Hole}, Ann. Phys.,82, 48-56
\bibitem[Wald,84]{R.Wald.84}Wald R.M., 1984
 \emph{General Relativity}, University of Chicago Press    
\bibitem[Wald,88]{Wald88}Wald R.M., 1988	
 \emph{Black Hole Thermodynamics} in \emph{Highlights in
 gravitation and cosmology}, Cambridge Univ. Press, Cambridge
\bibitem[Weinberg,72]{Weinberg}Weinberg S., 1972	
 \emph{Gravitation and Cosmology: Principles and Applications of
 the General Theory of Relativity}, Wiley, 165-170 
\bibitem[Witten,78]{Witten78}Witten Ed, 1978	
 \emph{}, Phys.Lett. {\bf B77}, 394 
\bibitem[Witten,88]{Witten88}Witten Ed, 1988
 \emph{}, Nucl.Phys. {\bf B311}, 46
\bibitem[Witten,95]{Witten952}Witten Ed, 1995
 \emph{}, Nucl.Phys. {\bf B443}, 85  
\bibitem[Witten,95,10]{Witten95}Witten Ed, 1995
 \emph{Bound States Of Strings And p-Branes}, Nucl.Phys. {\bf B460}, hep-th/9510135 
\bibitem[Witten,97]{Witten97}Witten Ed, 1997
 \emph{}, hep-th/9703166  
\bibitem[Youm,97]{Youm976}Youm D., 1997
 \emph{Entropy of Non-Extreme Rotating Black Holes in String Theories},
  hep-th/9706046
\end{thebibliography}
\end{document}